\documentclass[onecolumn,floatfix,superscriptaddress,showpacs,showkeys,nofootinbib,preprint]{revtex4-1}

\usepackage{amsmath}
\usepackage{mathtools}
\usepackage{amssymb}
\usepackage[english]{babel}
\usepackage[bookmarks=false,colorlinks=false]{hyperref}
\usepackage{graphicx}
\usepackage{color}
\usepackage{url}
\usepackage{footnote}
\usepackage{amsmath}
\usepackage[normalem]{ulem}

\renewcommand\sout{\bgroup \color{red}\ULdepth=-.5ex \ULset}

\RequirePackage{lineno}

\begin{document}

\title{The imprint of conservation laws on correlated particle production}

\author{P.~Braun-Munzinger}
\affiliation{Extreme Matter Institute EMMI, GSI Helmholtzzentrum f{\"u}r Schwerionenforschung, Darmstadt, Germany}
\affiliation{Physikalisches Institut, Universit\"{a}t Heidelberg, Heidelberg, Germany}

\author{K.~Redlich}
\affiliation{Institute of Theoretical Physics, University of Wroclaw, Wroclaw, Poland}

\author{A.~Rustamov}
\affiliation{GSI Helmholtzzentrum f{\"u}r Schwerionenforschung, Darmstadt, Germany} 
\affiliation{National Nuclear Research Center, Baku, Azerbaijan}

\author{J.~Stachel}
\affiliation{Physikalisches Institut, Universit\"{a}t Heidelberg, Heidelberg, Germany}

\begin{abstract}
The study of event-by-event fluctuations of net-baryon number in a subspace of full phase space is a promising direction for deciphering the structure of strongly interacting matter created in head-on collisions of relativistic heavy nuclei. Such fluctuations are generally suppressed by exact baryon number conservation. Moreover, the suppression is stronger if baryon number is conserved locally. In this report we present a conceptually new approach to quantify correlations in rapidity space between baryon-antibaryon, baryon-baryon, and antibaryon-antibaryon pairs and demonstrate their impact on net-baryon number fluctuations. For the special case of Gaussian rapidity distributions, we use the Cholesky factorization of the covariance matrix, while the general case is introduced by exploiting the well-known Metropolis and Simulated Annealing methods. The approach is based on the use of the canonical ensemble of statistical mechanics for baryon number and can be applied to study correlations between baryons as well as strange and/or charm hadrons. It can also be applied to describe relativistic nuclear collisions leading to the production of multi-particle final states. One application of our method is the search for formation of proton clusters at low collision energies emerging as a harbinger of the anticipated first-order chiral phase transition.  In a first step, the results obtained are compared to the recent measurements from the CERN ALICE  collaboration. Such investigations are key to explore the phase diagram of strongly interacting matter and baryon production mechanisms at energy scales from several GeV to several TeV. 
\end{abstract}
\maketitle

\section{Introduction} 
One of the key goals of nuclear collision experiments at high energy is to map the phase diagram of strongly interacting matter~\cite{Braun-Munzinger:2022bkc, Gross:2022hyw}.  Important first steps in this direction have been taken by studying first moments of the multiplicity distributions of produced particles in such collisions as function of collision energy, see  ~\cite{HRG,Andronic:2005yp,Andronic:2018qqt} and refs. given there. This has led to an experimental determination of the transition temperature between the hadronic and the partonic phase of QCD matter. 
A full determination of the QCD phase structure should include  measurements of higher-order moments of particle number distributions to search for critical behavior at or near the QCD phase boundary. At lower collision energies corresponding to values of baryon chemical potential $\mu_B > $ 400 MeV a critical end point of a first order chiral phase transition line might exist, see~\cite{Asakawa:1989bq, Rajagopal:1992qz, Stephanov:1998dy,Fu:2019hdw}. A promising tool to probe the critical phenomena is the study of fluctuations of conserved charges such as net-baryon number since, in a thermal system, such fluctuations are directly related to the equation of state (EoS) of the system under the study~\cite{Gross:2022hyw}.  One can  probe critical phenomena also at vanishing baryon chemical potential~\cite{LQCD1, Redlich1}. While at physical u,d,s-quark masses a rapid crossover chiral transition is found in Lattice QCD (LQCD) at pseudo-critical temperature $T_{pc} \approx$ 157 MeV~\cite{HotQCD:2018pds,Borsanyi:2020fev}, its proximity to a genuine 2nd order phase transition of O(4) universality class \cite{Pisarski:1983ms}, realised for vanishing u,d- quark masses, suggests remnants of criticality \cite{Ejiri:2005wq}. On the other hand, the pseudo-critical temperature, reported from LQCD~\cite{HotQCD:2018pds,Borsanyi:2020fev}, is in close agreement with the chemical freeze-out temperature as extracted by comparing  Hadron Resonance Gas (HRG) model predictions~\cite{Andronic:2005yp,HRG,Andronic:2018qqt} to the measured hadron multiplicities for collision energies investigated at the SPS, RHIC and LHC accelerators. This agreement implies that strongly interacting matter, created in collisions of heavy nuclei at these energies, freezes out in close vicinity of the chiral phase transition line. Consequently, the effect of singularities stemming from the proximity to a second order chiral phase transition can be investigated via fluctuation measurements at small or vanishing net-baryon densities. This offers unique possibilities for verifying LQCD predictions for the crossover transition.

The current net-proton number fluctuation measurements by the HADES collaboration at SIS18~\cite{HADES:2020wpc}, by the STAR collaboration at RHIC \cite{STAR:2020tga, STAR:2021fge, STAR:2021rls, STAR:2021iop,Xu:2018vnf,STAR:2022etb}, and by ALICE at the LHC \cite{RustamovQM17,ALICE:2019nbs,ALICE:2022xpf}, have provided interesting and stimulating results. However, no direct evidence of criticality has emerged yet, and quantitative analysis of the measurements is made difficult by the presence of non-critical effects such as volume or participant fluctuations \cite{Gorenstein:2011vq, Skokov:2012ds, Braun-Munzinger:2016yjz, Rustamov:2022sqm} and by correlations introduced by baryon number conservation~\cite{Bzdak:2012an, Braun-Munzinger:2016yjz, Braun-Munzinger:2018yru, Braun-Munzinger:2020jbk}. 
In statistical mechanics, fluctuations of conserved quantities, such as the baryon number, are described within the Grand Canonical Ensemble (GCE)~\cite{StatLandau} formulation, where only the average values of these quantities are conserved. In general, fluctuation measurements are performed over a fraction of total phase space. Contrary to the GCE calculations, the results of measurements, even in a limited acceptance, are modified by exact conservation of charges in full phase space. To justify the comparison of theoretical calculations within a GCE, such as the HRG~\cite{HRG} and LQCD~\cite{LQCD1}, to experimental results, the effects of conservation laws on experimental data need to be accounted for~\cite{Bzdak:2012an, Braun-Munzinger:2018yru, Braun-Munzinger:2020jbk, Vovchenko:2022szk,Poberezhnyuk:2022blc}. Analyses of net-baryon number fluctuations with emphasis on conservation laws have been presented in ~\cite{ALICE:2019nbs,ALICE:2022xpf}.

Here, in the framework of the Canonical Ensemble (CE) formulation, we provide first quantitative estimates of the implications of local baryon number conservation for measurements in a finite acceptance. First estimates for the importance of local baryon number conservation are presented in ~\cite{Braun-Munzinger:2019yxj}. 

In the following, we first provide a summary of conservation laws in the CE and discuss when and  where it is applicable. We then briefly review the connection between local baryon number conservation and the width in rapidity of the fireball formed in the collision. How to deal with correlations other than those induced by conservation laws is one of the major new aspects treated here. Finally, we will discuss applications of the new fluctuation analysis proposed here using the example of data from the LHC.

\section{Conservation laws via canonical thermodynamics}

Exact conservation laws such as baryon number conservation can be imposed by using the CE formulation of statistical mechanics~\cite{Redlich:1979bf, Hagedorn:1984uy,Braun-Munzinger:2003pwq,Begun:2004zb,Becattini:1995if}. This has been described in detail in ~\cite{Braun-Munzinger:2020jbk, Bzdak:2012an}.
For a thermal system with  volume $V$ at temperature $T$, composed of particles with baryon numbers $b=\pm 1$ and total net-baryon number $B$, the CE partition function for baryons reads:

\begin{equation}
Z^{CE}(B, T, V) = \frac{1}{2\pi}\int_0^{2\pi}d\phi \,e^{-iB\phi} exp\left[ Z^{b}_{+1}e^{i\phi}+Z^{b}_{-1}e^{-i\phi}\right] = \left(\frac{Z^{b}_{+1}}{Z^{b}_{-1}}\right)^{\frac{B}{2}}I_{B}\left(2\sqrt{Z^{b}_{+1}Z^{b}_{-1}}\right),
\label{bcan}
\end{equation}
where $I_{B}$ is the modified Bessel function of the first kind \cite{Abramowitz:1972}, $Z^{b}_{+1}$ and $Z^{b}_{-1}$ denote the sum of single-particle partition functions for all particles having baryon numbers $+1$ and $-1$, respectively:

\begin{equation}
Z^{b}_{\pm 1}=\sum_{i}z_{i,b=\pm 1},
\label{singlepart}
\end{equation}
with the single-particle partition function for a baryon with mass $m_{i}$, strangeness $S_{i}$, electric charge $Q_{i}$ and degeneracy factor $g_{i}$ defined as:

\begin{equation}
z_{i} = \frac{Vg_{i}}{(2\pi)^{3}}\int d^{3}p \, exp[(-E_{i}+S_{i}\mu_{S}+Q_{i}\mu_{Q})/T],
\end{equation}
where $\mu_{S}$ and $\mu_{Q}$ denote the strangeness and electric charge chemical potentials and $E_{i} = \sqrt{p^{2}+m_{i}^{2}}$.

Mean multiplicities of baryons can be obtained from the partition function in Eq.~\ref{bcan} by introducing fictitious fugacities ($\lambda_{i}$) in Eq.~\ref{singlepart}, such that $z_{i} \rightarrow \lambda_{i}z_{i}$

\begin{equation}
\langle N_{i}\rangle^{CE}_{b=+1} = \left.\lambda_{i}\frac{\partial lnZ^{CE}}{\partial \lambda_{i}}\right |_{\lambda_{i}=1}= z_{i}\sqrt{\frac{Z^b_{-1}}{Z^b_{+1}}} \frac{I_{1}\left(2\sqrt{Z^{b}_{+1}Z^{b}_{-1}}\right)}{I_{0}\left(2\sqrt{Z^{b}_{+1}Z^{b}_{-1}}\right)}.
\label{multce1}
\end{equation}

The canonical volume $V_{C}$ is contained in the argument of the Bessel functions, inside the $Z^{b}_{+1}$ and $Z^{b}_{-1}$. Since we focus here on data from the LHC we consider the high energy limit, i.e. we assume that $B$ = 0 and all chemical potentials vanish (which leads to $Z^{b}_{+1} = Z^{b}_{-1} = Z^{b}$), and explicitly demonstrate the appearance of the canonical volume $V_{C}$:

\begin{equation}
\langle N_{i}\rangle_{b=+1} = V\langle n_{i}\rangle \frac{I_{1}(2Z^{b})}{I_{0}(2Z^{b})} = V\langle n_{i}\rangle\frac{I_{1}(2V_{c}\sum_{i} \langle n_{i}\rangle_{b=+1})}{I_{0}(2V_{c}\sum_{i} \langle n_{i}\rangle_{b=+1})},
\label{multce2}
\end{equation}
with $\langle n_{i}\rangle$ standing for Grand Canonical densities of particles.

The volume $V$ entering Eq.~\ref{multce2} is determined by confronting experimentally measured charged particle yields with those calculated in thermal models within the GCE; at LHC energy it typically corresponds to the fraction of the overall volume determined by particle yields measured within one unit of rapidity around midrapidity. The ratio $I_{1}(2Z^{b})/I_{0}(2Z^{b})$ is called a canonical suppression factor~\cite{Cleymans:1996yg, Cleymans:1990mn, Hagedorn:1984uy,Hamieh:2000tk,Cleymans:1998yb,Braun-Munzinger:2003pwq}.
For large values of $Z^{b}$ the suppression factor approaches unity and one recovers GCE multiplicities~\cite{Tounsi:2001ck,Sharma:2022poi}. In fact, there are no fundamental reasons to introduce two volumes, $V$ and $V_{c}$. However, historically, $V_{c}$ was  introduced as an additional parameter, independent of $V$. The required decrease of $Z^{b}$ is then, as seen from Eq.~\ref{multce2}, due to the introduction of $V_{C}$. Hence, if the predicted particle yield at a given temperature is higher than the experimental measurements, one can apply  a canonical suppression factor by selecting  a smaller value for the volume  $V_{C}$. This is the essential idea of introducing conservation laws via a canonical volume $V_{c}$. The disadvantage of this approach is that the introduced volume $V_{c}$ does not correspond to a physical quantity. As mentioned above, it is just an additional parameter of the model which can but does not have to be smaller than the volume $V$. This approach does however not allow for a simultaneous treatment of global and local conservation laws.

 As a recent example of the use of canonical thermodynamics to establish a non-critical baseline for fluctuation measurements we refer to ~\cite{Braun-Munzinger:2020jbk}. Fluctuations of baryon number appear in a fraction of the overall volume, while in the overall volume baryon number does not fluctuate. The amount of fluctuations, however, depends on possible more local conservation laws, such as correlations between baryons and anti-baryons in rapidity space. If the correlation is strong and, accordingly, the correlation length is short, the probability of having both baryons and anti-baryons inside a given sub-volume increases, thereby suppressing the amount of net-baryon fluctuations. Within this scheme particles can be created anywhere in the available rapidity space, however the corresponding anti-particle production depends on the correlation length. A large (infinite) correlation length corresponds to a global conservation law, whereas a finite correlation length accounts for local conservation effects. 
 Here we consider Pb--Pb collisions at the LHC. At or close to the current top energy of $\sqrt{s_{NN}}$= 5.02 TeV, the two Pb beams travel before the collision at rapidities $y = \pm 8.6$. In a head-on collision between two Pb nuclei, the participant nucleons undergo a significant rapidity shift of about $\delta y = 2$ ~\cite{Braun-Munzinger:2020jbk}, implying that the participants end up in fragmentation regions at rapidities $y \approx \pm 6.6$ with an estimated width of about one unit of rapidity. This is based on the observation that the concept of limiting fragmentation is well describing at intermediate RHIC energies where a measurement is still possible ~\cite{BRAHMS:2009wlg}. Each fragmentation region carries the large baryon and electric-charge number of the corresponding beam and is quite well separated from the central region of particle production in which the bulk of produced particles is located. From the ALICE measurement of the multiplicity of produced charged particles in such a collision ~\cite{ALICE:2016fbt} one can estimate that the fireball of produced particles centered at $y = 0$ extends to about $\pm 5.5$ units of rapidity, implying that this fireball has very small net-baryon and net-charge number, but contains the energy deposited by the beams in the central fireball. In particular, the central region near $|y| < 1$ should contain an equal number of baryons and anti-baryons. For a recent experimental demonstration of this, see ~\cite{ALICE:2023ulv}; the proton number exceeds the antiproton number by only about 1 \%. One can estimate ~\cite{Braun-Munzinger:2014lba} the volume of the slice  at rapidity 0 and $\delta y = \pm 0.5$ by using Bjorken's concept of space-time rapidity and by making use of the the ALICE pion HBT measurement ~\cite{ALICE:2011dyt} to yield $V_0 \approx 6900$ fm$^3$ at full LHC energy. The total central fireball volume $V$ is the volume of $\pm 5.5$ units of rapidity. Only within this huge volume baryon number and electric charge must be strictly conserved.

 As the novel idea of the current article is to account for correlated baryon-antibaryon production, we proceed further by introducing methods for generating correlated random numbers.

\section{Correlated random numbers}

\subsection{Gaussian rapidity distributions}
\label{Cholesky}
 The generation of correlated random numbers plays an important role in many studies in nuclear and particle physics. To date, no general analytical procedure is known for introducing correlations between random numbers drawn from arbitrary distributions. However, for random numbers drawn from Gaussian distributions, the Cholesky decomposition (or factorization) ~\cite{Cholesky:1996ab} can be exploited. The Cholesky decomposition allows to transform uncorrelated standard normal variables into correlated normal variables by multiplying them with a lower triangular matrix derived from the covariance matrix.  In the following, we demonstrate this method to generate pairs of random variables  $x_{1}$, $x_{2}$ with the desired correlation coefficient $\rho$. We first decompose the co-variance matrix $\Sigma$ into a lower triangular matrix $L$, such that $LL^{T}=\Sigma$, where $L^{T}$ is a transpose matrix of $L$ 

\begin{equation}
\label{defCorrMatrix}
\Sigma = 
\begin{pmatrix}
\sigma_{x_{1}}^{2} & \rho\sigma_{x_{1}}\sigma_{x_{2}} \\
\rho\sigma_{x_{1}}\sigma_{x_{2}} &\sigma_{x_{2}}^{2} 
\end{pmatrix} = LL^{T},
\end{equation}
from where it follows that

\begin{equation}
L = 
\begin{pmatrix}
\sigma_{x_{1}} & 0\\
\rho\sigma_{x_{2}}&\sqrt{1-\rho^{2}}\sigma_{x_{2}}.
\end{pmatrix}.
\end{equation}

The generation of correlated pairs of random numbers is then done in two steps:\\
i) Two uncorrelated random numbers $y_{1}$ and $y_{2}$ are generated from a standard normal distribution, i.e., with $\langle y1\rangle=\langle y_{2}\rangle = 0$ and $\sigma_{y1}=\sigma_{y_{2}} = 1$.\\
ii) Correlated variables $x_{1}$ and $x_{2}$ are then obtained as 

\begin{equation}
\begin{pmatrix}
x_{1}  \\
x_{2} 
\end{pmatrix} = 
 \begin{pmatrix}
\langle x_{1} \rangle  \\
\langle x_{2} \rangle 
\end{pmatrix}+
L \begin{pmatrix}
y_{1}  \\
y_{2}
\end{pmatrix}.
\end{equation}

It is rather straightforward to verify that the so generated set of $x_{1}$ and $x_{2}$ values are normally distributed and correlated with the corresponding correlation coefficient $\rho$ entering into the co-variance matrix given in Eq.~\ref{defCorrMatrix}.

\subsection{Arbitrary rapidity distributions}
\label{sec_metropolis}

In actual collisions of nuclei at high energies, the measured rapidity distributions of particles, in general, do not follow normal  distributions. This is already the case at top AGS energy with $\sqrt{s_{NN}} \approx 5$ GeV ~\cite{E877:1999qdc,E-802:1998xum}. At very high energy $\sqrt{s_{NN}} > 5$ TeV the rapidity distribution of protons exhibits structures caused by the beam and target fragmentation as well as a central region (see above). The shape of the central region is given by temperature and longitudinal expansion of the fireball. Thus it becomes necessary to generate randomly correlated numbers with arbitrary probability distribution functions. For this general case we exploit the well known Metropolis ~\cite{Metropolis:1953am, Charmpis:2004} and Simulated Annealing ~\cite{Kirkpatrick:1983zz}  algorithms. Our starting point are rapidity distributions of baryons and anti-baryons; we randomly generate  rapidity values for baryons and anti-baryons in two sets $\{y_{1}^{B},y_{2}^{B}, y_{3}^{B}, ... \}$ and $\{y_{1}^{\bar{B}},y_{2}^{\bar{B}}, y_{3}^{\bar{B}}, ... \}$. At this point there are no correlations between the two sets which corresponds to a vanishing correlation coefficient

\begin{equation}
\rho=\frac{Cov(y_{B},y_{\bar{B}})}{\sigma_{B}\sigma_{\bar{B}}},
\label{corrcoeff}
\end{equation}

where $Cov(y_{B},y_{\bar{B}})$ is the covariance and $\sigma_{B}$ is the standard deviation, defined as

\begin{equation}
Cov(y_{B},y_{\bar{B}})=\frac{\sum_{i=1}^{N}\left(y_{i}^{B}-\langle y_{B}\rangle\right)\left(y_{i}^{\bar{B}}-\langle y_{\bar{B}}\rangle\right)}{N-1} = \langle y_{B}y_{\bar{B}}\rangle - \langle y_{B}\rangle \langle y_{\bar{B}}\rangle,
\end{equation}

\begin{equation}
\sigma_{B}^{2}=\frac{\sum_{i=1}^{N}\left(y_{i}^{B}-\langle y_{B}\rangle\right)^{2}}{N-1}= \langle y_{B}^{2} \rangle - \langle y_{B} \rangle^{2}.
\end{equation}

\begin{figure}[htb]
\centering
 \includegraphics[width=1.\linewidth,clip=true]{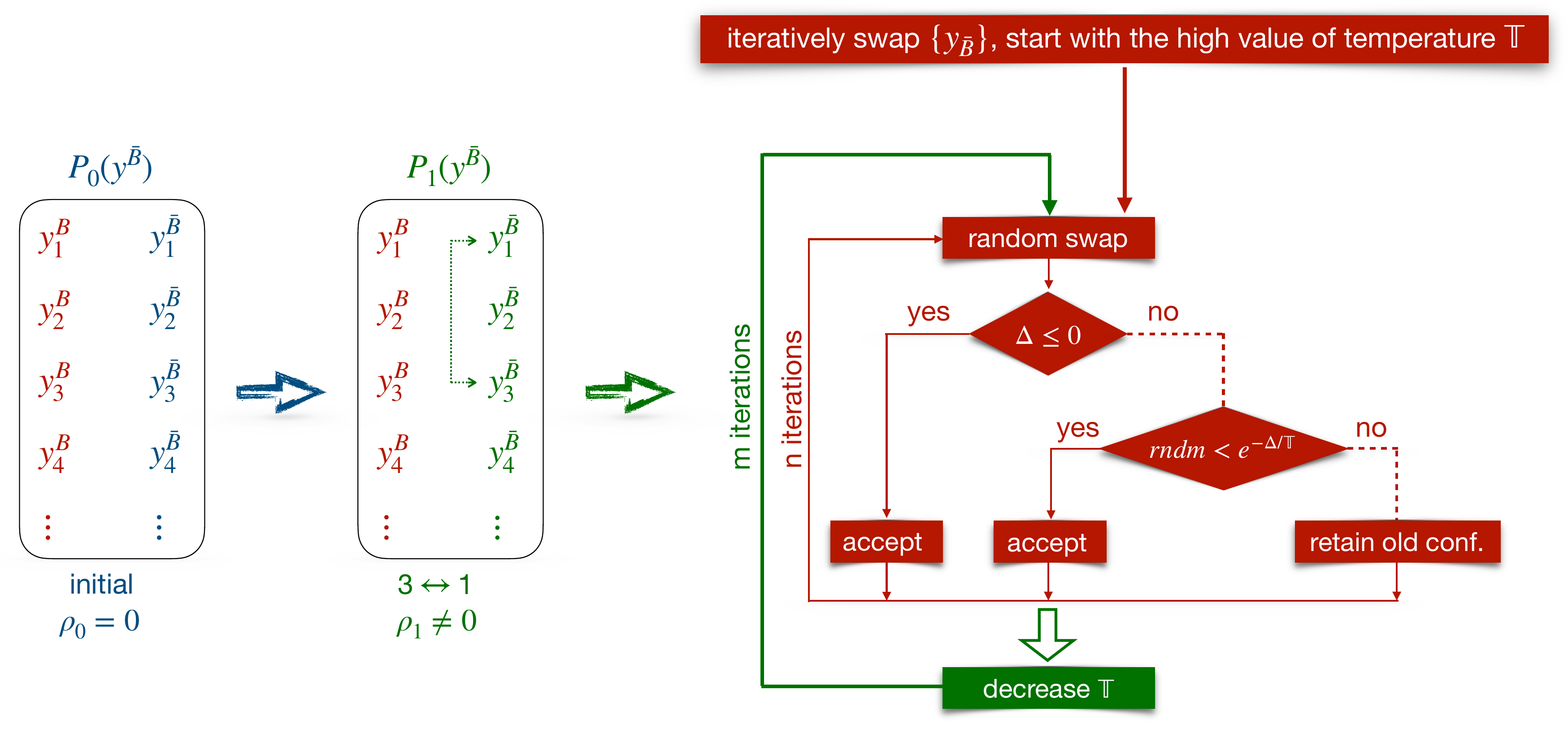}
 \caption{Implementation of the Metropolis algorithm.}
\label{fig_metropolis}
\end{figure}

Next, to introduce correlations in rapidity space between baryons and antibaryons, we randomly swap the rapidity values for antibaryons, thus generating an updated set of rapidity values for antibaryons, while the rapidities of baryons are kept unmodified. This is schematically illustrated in the left part of Fig.~\ref{fig_metropolis}. 
The acceptance rate for the updated set is based on the Metropolis algorithm ~\cite{Metropolis:1953am, Charmpis:2004} and obtained in two loops as illustrated in the right part of Fig.~\ref{fig_metropolis}. The inner loop consists of $n$ iterations, the outer loop of $m$ iterations. The goal of the inner loop is to produce a thermalized system at a temperature $\mathbb{T}$. Within the inner loop at each iteration $n$ the cost function $\Delta$ is computed

\begin{equation}
\Delta = |\rho_{n}-\rho| - |\rho_{n-1}-\rho|,
\end{equation}
where $\rho$ denotes the "desired" correlation coefficient and $\rho_{n-1}$ and $\rho_{n}$ are the values of the correlation coefficients as computed using Eq.~\ref{corrcoeff} for two successive sets of rapidities for antibaryons. If $\Delta$ $\le 0$ the swapped set is accepted. 
On the other hand, if $\Delta >0$, the acceptance  probability is taken to be  $p=exp[-\Delta/\mathbb{T}]$. After completing the inner loop, the value of the temperature $\mathbb{T}$ is reduced.
For each inner iteration the percentage of the accepted swaps is calculated. The algorithm terminates after $m$ outer iterations when the percentage of the accepted swaps drops below 1 $\%$, that is when the global solution to the problem is reached. The number of inner iterations $n$ depends on the number of the initially generated uncorrelated rapidity values. In our case we generated $N$ = 200 rapidity values for baryons and antibaryons in each event and took $n$ = 5$\times N$. The initial temperature value is set to 0.01 and it is decreased in each outer loop by a factor of 0.85. This ansatz is based on Ref.~\cite{Charmpis:2004} and was further optimized for our studies. 
The result of this algorithm is presented in the left panel of Fig.~\ref{fig_corr_length}, which represents a correlated rapidity distributions with the correlation coefficient $\rho$ = 0.6. The procedure also allows to introduce anti-correlations. 

\subsection{Correlation length}

\begin{figure}[htb]
\centering
 \includegraphics[width=1.\linewidth,clip=true]{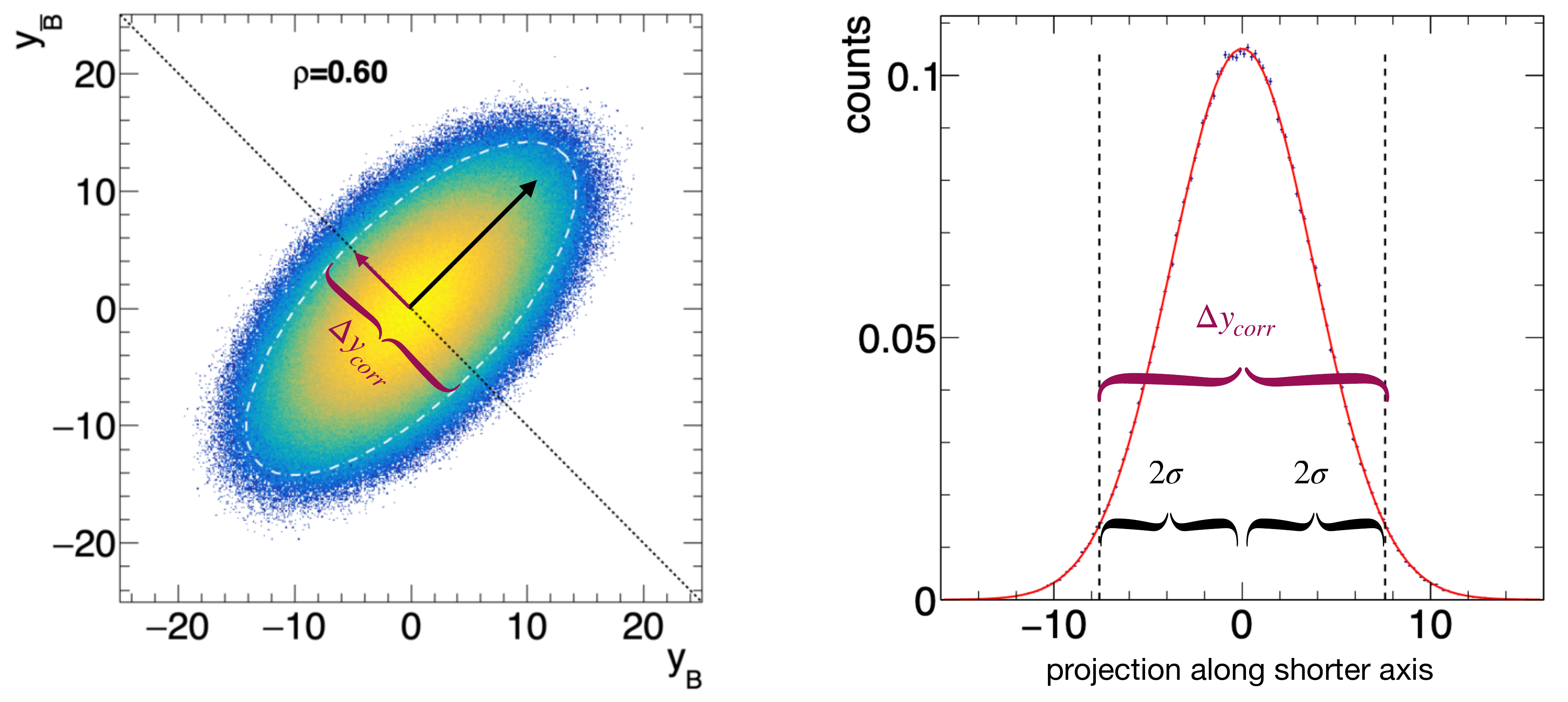}
 \caption{Correlated baryon and antibaryon rapidity distribution for a correlation coefficient $\rho$ = 0.6 (left) as well as it's projection on the shorter axis (right).}
\label{fig_corr_length}
\end{figure}

It is more intuitive to quantify the correlation in rapidity space with the correlation length $\Delta y_{corr}$ instead of the correlation coefficient $\rho$. The relation between the two definitions is illustrated in the left panel of Fig.~\ref{fig_corr_length}. In the first step the obtained 2D distribution is projected onto the shorter axis of the covariance ellipse. The so obtained projection plot is illustrated in the right panel of Fig.~\ref{fig_corr_length}. Next the $\sigma$ of this distribution is computed and $\Delta y_{corr}$ is taken as $\pm 2\sigma$. Thus the correlations in the rapidity space can be quantified with either the correlation coefficient $\rho$ or the corresponding correlation length $\Delta y_{corr}$. 

The relation between $\rho$  and $\Delta y_{corr}$ can also be obtained analytically by solving the eigen-equation for the covariance matrix:

\begin{equation}
\Sigma \vec{v} = \lambda \vec{v},
\label{Eq_EigenEq}
\end{equation}
where $\vec{v}$, and $\lambda$ denote the eigenvector and eigenvalue of the covariance matrix $\Sigma$:

\begin{equation}
\label{defCorrMatrix}
\Sigma = 
\begin{pmatrix}
\sigma_{B}^{2} & \rho\sigma_{B}\sigma_{\bar{B}} \\
\rho\sigma_{B}\sigma_{\bar{B}} &\sigma_{\bar{B}}^{2} 
\end{pmatrix}.
\end{equation}

The straightforward solution of Eq.\ref{Eq_EigenEq} yields the following two solutions for the eigenvalues:

\begin{equation}
\begin{pmatrix}
\lambda_{1}  \\
\lambda_{2} 
\end{pmatrix} = 
\begin{pmatrix}
\frac{\sigma_{B}^{2}+\sigma_{\bar{B}}^{2}}{2}- \frac{1}{2}\sqrt{2\sigma_{B}^{2}\sigma_{\bar{B}}^{2}(2\rho^{2}-1)+\sigma_{B}^{4}+\sigma_{\bar{B}}^{4}} \\
\frac{\sigma_{B}^{2}+\sigma_{\bar{B}}^{2}}{2}+ \frac{1}{2}\sqrt{2\sigma_{B}^{2}\sigma_{\bar{B}}^{2}(2\rho^{2}-1)+\sigma_{B}^{4}+\sigma_{\bar{B}}^{4}} \\
\end{pmatrix},
\label{Eq_EigenValues}
\end{equation}
where, $\lambda_{1}$ and $\lambda_{2}$ represent the variance of the distribution along the shorter and longer axes of the covariance ellipse, respectively. 

The  eigenvectors $\vec{v}$ of the covariance matrix $\Sigma$, representing  orientations of the shorter and longer axes of the covariance ellipse can be obtained via substituting the eigenvalues from Eq.~\ref{Eq_EigenValues} into Eq.~\ref{Eq_EigenEq}. These eigenvectors are not needed in the current work.

We note that, for the special case of $\sigma_{B} = \sigma_{\bar{B}}$, which is valid at LHC energies, the eigenvectors coincide with the diagonals (cf. Fig.~\ref{fig_corr_length}), and  the correlation length can be calculated as:

\begin{equation}
\Delta y_{corr} \equiv 4\sqrt{\lambda_{1}} = 4\sigma_{B}\sqrt{1-\rho}.
\label{Eq_coeffVsCorr1}
\end{equation}

\subsection{Canonical Ensemble with the Metropolis Algorithm}

\begin{figure}[htb]
\centering
 \includegraphics[width=0.6\linewidth,clip=true]{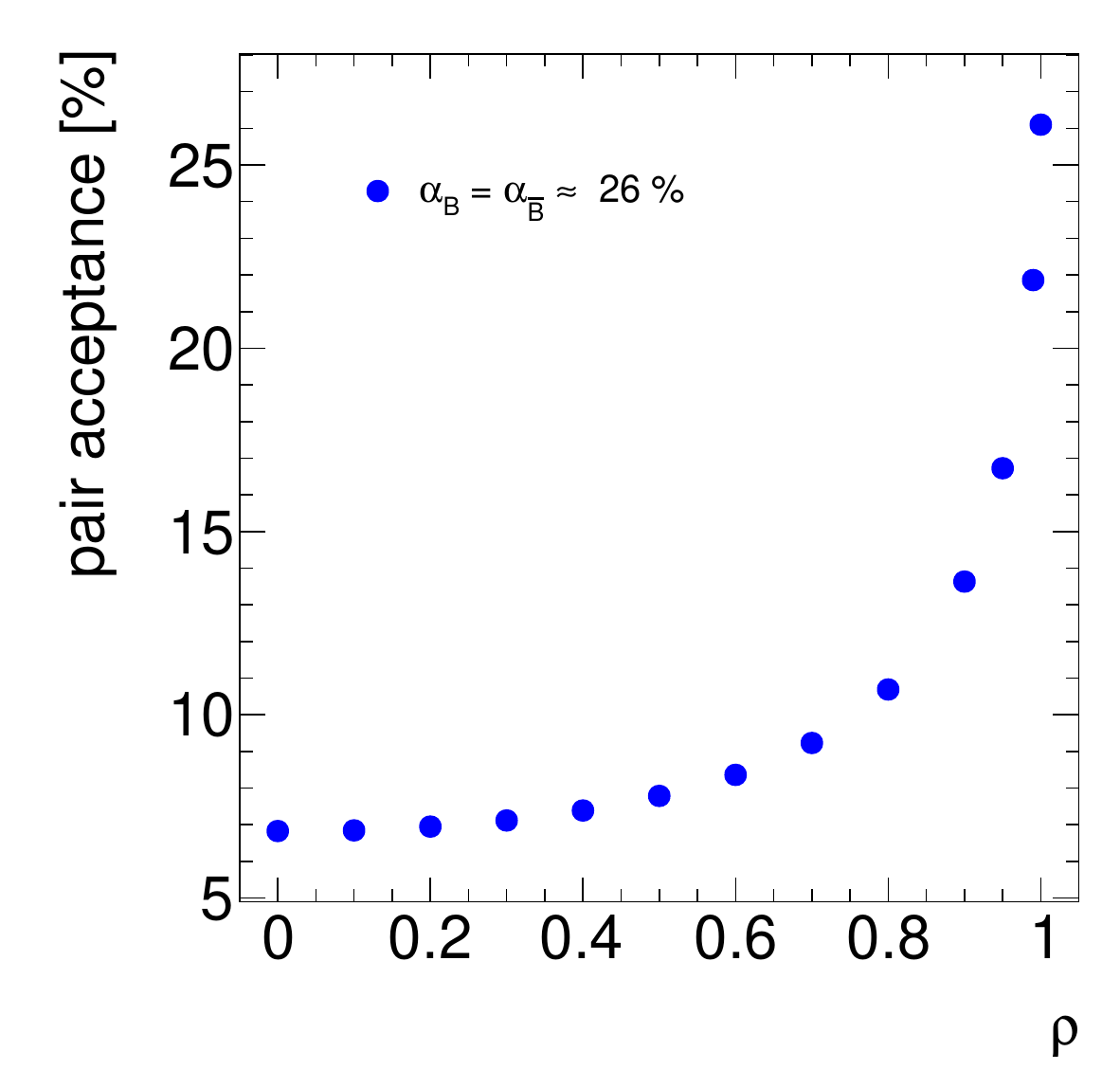}
 \caption{Acceptance of baryon-antibaryon pairs as a function of the correlation coefficient $\rho$ within a fixed acceptance of 26 \% for single baryons. For a vanishing value of the correlation coefficient the pair acceptance equals to the square of the single baryon acceptance, gradually approaching the single baryon acceptance as $\rho$ approaches 1.}
\label{acc_pair}
\end{figure}

\begin{figure}[htb]
\centering
 \includegraphics[width=0.45\linewidth,clip=true]{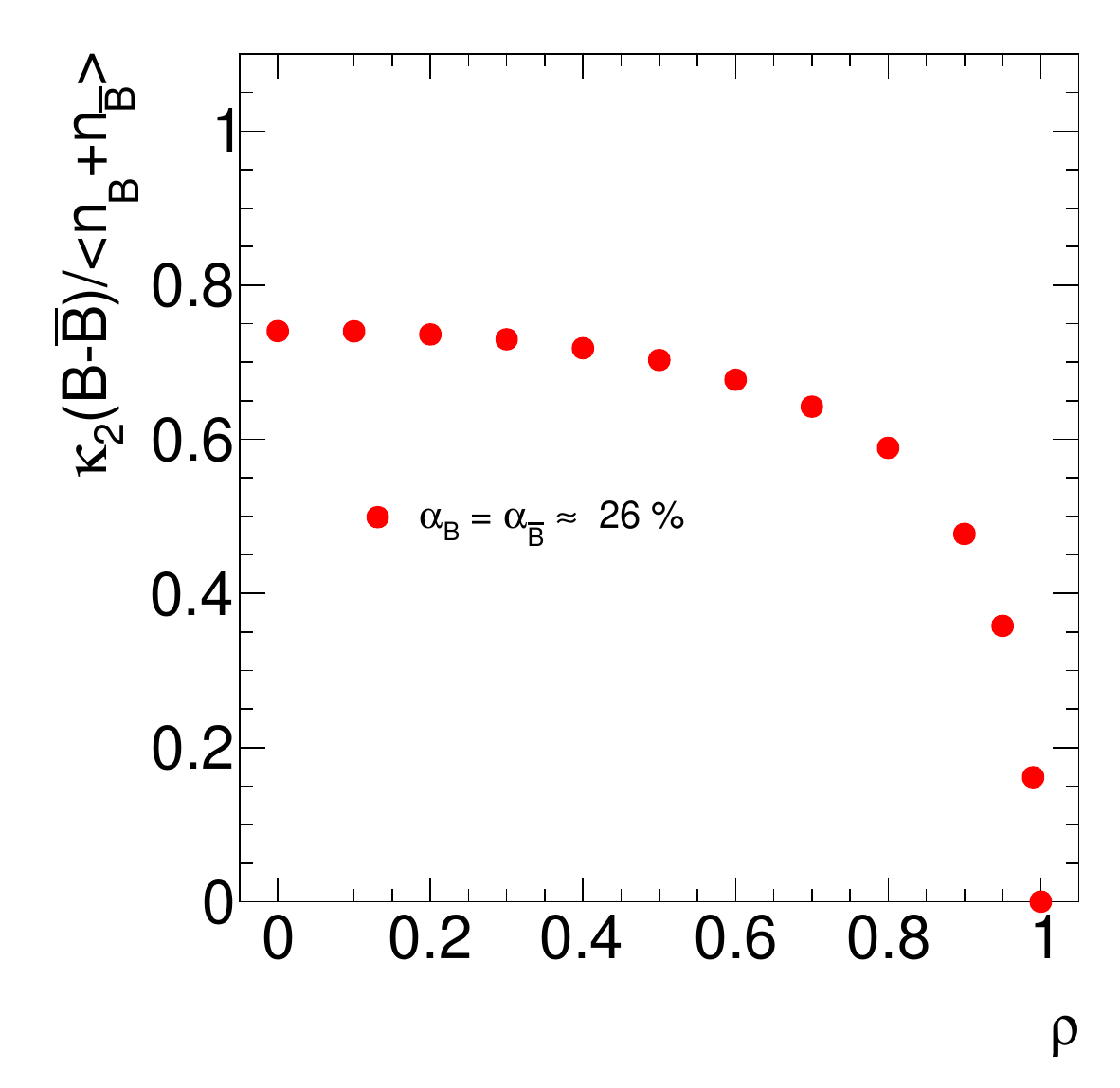}
  \includegraphics[width=0.45\linewidth,clip=true]{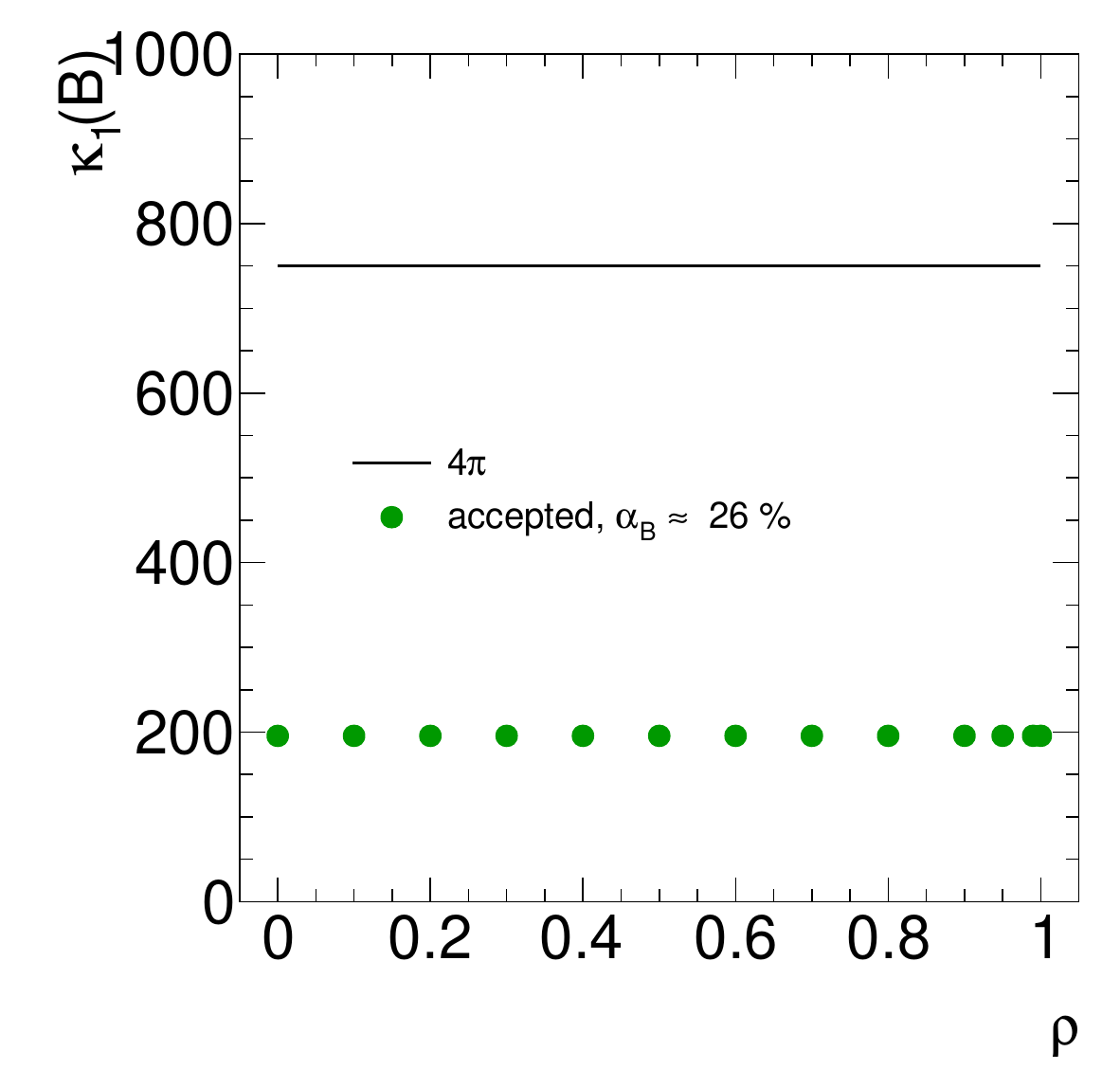}
 \caption{Left panel: Second order cumulants of net-baryons inside a finite acceptance of 26~\% normalized to the mean number of baryons plus antibaryons as a function of the correlation coefficient $\rho$. Right panel: First order cumulants, i.e. mean values of the number of baryons, inside the finite acceptance as a function of correlation coefficient (green symbols). The mean number of baryons in full phase space is depicted as the black line.}
\label{kappavsrho}
\end{figure}

In this section we include correlations between baryons and antibaryons in the CE partition function~\cite{Braun-Munzinger:2020jbk}. The probability $P(n_{B},n_{\bar{B}})$ of having $n_{B}$ baryons and $n_{\bar{B}}$ antibaryons inside the acceptance depends both on the correlation coefficient $\rho$ (or the correlation length $\Delta y_{corr}$) and the acceptance cut $y_{cut}$. It is defined as: 

\begin{eqnarray}\label{eq:canonical-partition} 
P(n_{B},n_{\bar{B}}) = \sum_{N_{B}}\sum_{N_{\bar{B}}}P(N_{B},N_{\bar{B}})P(n_{B},n_{\bar{B}}; \rho, y_{cut}, N_{B}, N_{\bar{B}}),
\label{eq:zCE}
\end{eqnarray}
where $P(N_{B},N_{\bar{B}})$  is the probability of having $N_{B}$ baryons and $N_{\bar{B}}$ antibaryons in the full phase space. On the other hand, $P(n_{B},n_{\bar{B}}; \rho, y_{cut}, N_{B}, N_{\bar{B}})$ entering Eq.~\ref{eq:zCE} is the probability of having $n_{B}$ baryons and $n_{\bar{B}}$ antibaryons inside the acceptance, defined by  $y_{cut}$, given that in full phase space there are $N_{B}$ baryons and $N_{\bar{B}}$ antibaryons. This probability depends on the correlation coefficient between rapidities of baryons and antibaryons. The larger the correlation coefficient the bigger the probability of accepting baryons and anti-baryons in pairs, which in turn reduces the amount of fluctuations for net-baryons. This is because, in full phase space, net-baryon number does not fluctuate as it is treated within the CE. For example, using Gaussian rapidity distributions for baryons and anti-baryons and the Cholesky decomposition discussed in section~\ref{Cholesky}, we present in Fig.~\ref{acc_pair} the pair acceptance as a function of correlation length for a fixed value of $y_{cut}$ corresponding to $\alpha_{B}$ = 26 $\%$ acceptance for single baryons or antibaryons.  To demonstrate the effect of the correlation coefficient, we used as an example $N_{B}$ = $N_{\bar{B}}$ = 750 in full phase space. These values are close to the actual values for the central fireball in Pb-Pb collisions at $\sqrt{s_{NN}}$ = 2.76 TeV~\cite{Braun-Munzinger:2020jbk}. For $P(N_{B},N_{\bar{B}})$ we used the expression derived in our previous work~\cite{Braun-Munzinger:2020jbk}, where exact baryon number conservation is imposed by exploiting the canonical partition function for a thermal system composed of baryons and antibaryons with the HRG Equation of State (EoS). The corresponding probability distribution reads:

\begin{eqnarray}\label{prob}
P(N_{B}, N_{\bar{B}})=\frac{1}{I_{B}(2z)}\frac{z^{B}z^{2N_{\bar{B}}}}{(N_{\bar{B}}+B)!N_{\bar{B}}!},
\label{eq:nCE}
\end{eqnarray}
where the single-particle partition function $z=750.25$ is calculated using the mean multiplicities mentioned above (see also ~\cite{Braun-Munzinger:2020jbk}). In fact, Eq.~\ref{prob} is used to produce the number of antibaryons $N_{\bar{B}}$, while the number of baryons is fixed in each event  $N_{B} = N_{\bar{B}} + B$ (with $B = 0$ for our specific example). 
As seen in Fig.~\ref{acc_pair}, for $\rho$ = 0.0, i.e. when there is no correlation between baryons and antibaryons, the pair acceptance is $\alpha_{B}^{2}$, while for full correlation ($\rho$ = 1.0) it is equal to $\alpha_{B}=\alpha_{\bar{B}}$. Thus, with increasing correlation coefficient from zero to unity, net-baryon fluctuations progressively vanish. This is clearly visible in the left panel of Fig.~\ref{kappavsrho}, where the normalized second-order cumulants of net-baryons as a function of the correlation coefficient $\rho$ is presented. We note that the correlation length introduced here does not affect mean multiplicities of baryons within a given acceptance, as illustrated in the right panel of Fig.~\ref{kappavsrho}. Hence the correlation length can be obtained using only higher order moments of net-baryon distributions.

The procedure described above allows for the introduction of correlations between baryon-baryon, baryon-antibaryon and antibaryon-antibaryon pairs. Moreover, it is also possible to include correlations between multiple number of baryons, which maybe  important when searching for clustered production of baryons, emerging, e.g., from spinodal instabilities~\cite{Sasaki:2007db}.

\begin{figure}[htb]
\centering
 \includegraphics[width=1.\linewidth,clip=true]{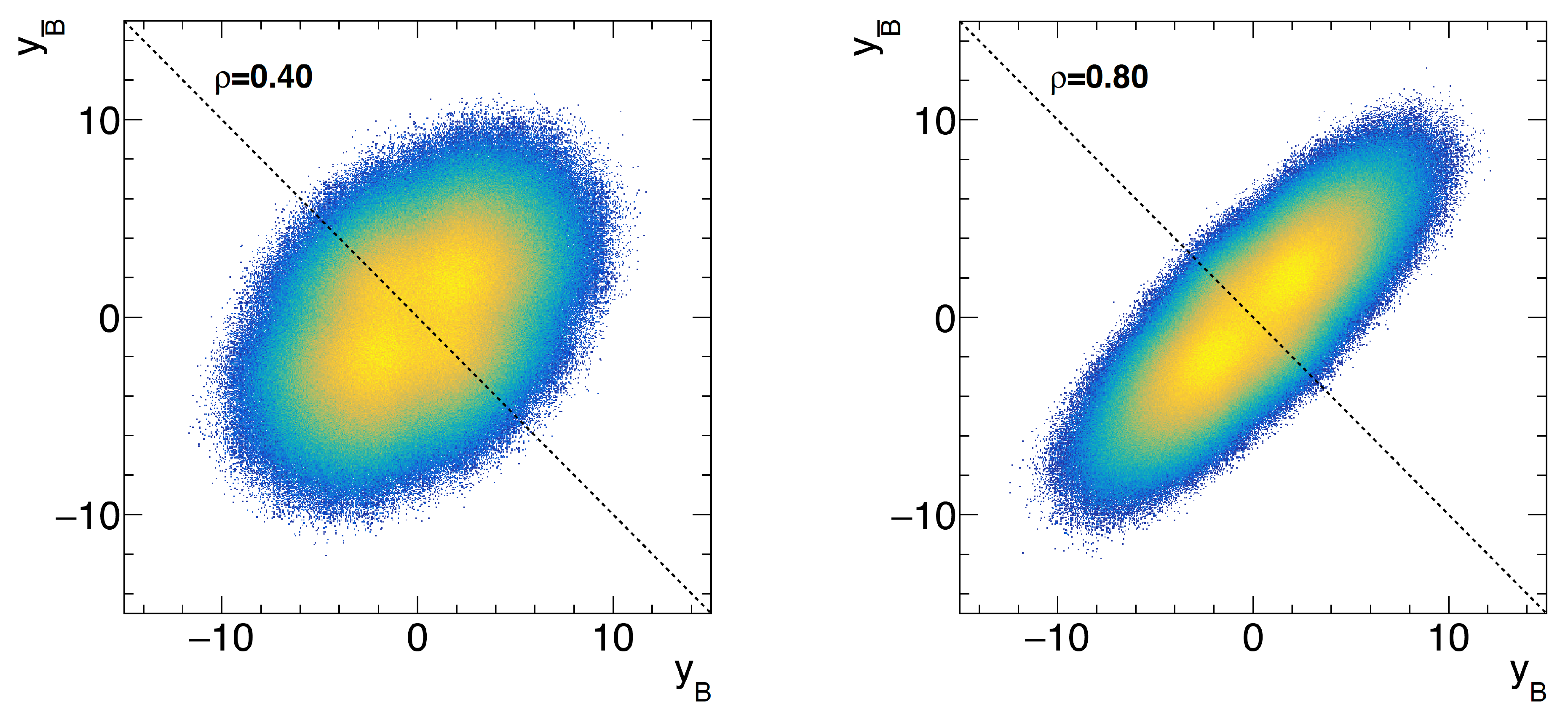}
 \caption{The correlation between rapidity values of baryons and anti-baryons for $\rho=0.4$ (Left panel) and $\rho=0.8$ (Right panel). The dashed diagonal lines show the directions orthogonal to the largest eigenvector of the corresponding co-variance matrices.} 
\label{ALICE04}
\end{figure} 

The only input information needed to generate the correlated particle production in our approach are rapidity distributions of baryons and antibaryons and the desired correlation length as well as the experimental acceptance $y_{cut}$. In the case of non-Gaussian rapidity distributions we use the Metropolis algorithm to obtain $P(n_{B},n_{\bar{B}}; \rho, y_{cut}, N_{B}, N_{\bar{B}})$ entering into Eq.~\ref{eq:zCE}. In the output, our approach provides the following information:\\ 
(i) the number of baryons and anti-baryons in each event,\\
(ii) the acceptance of baryons and anti-baryons, and\\
(iii) the correlation length. 

The so generated events are analysed and the obtained results on cumulants are confronted with the experimental measurements. The correlation coefficient is extracted, by fitting the experimental measurements of cumulants of net-baryons to the simulated ones within the CE, as described above. 

\section{Results}
\begin{figure}[htb]
\centering
 \includegraphics[width=0.45\linewidth,clip=true]{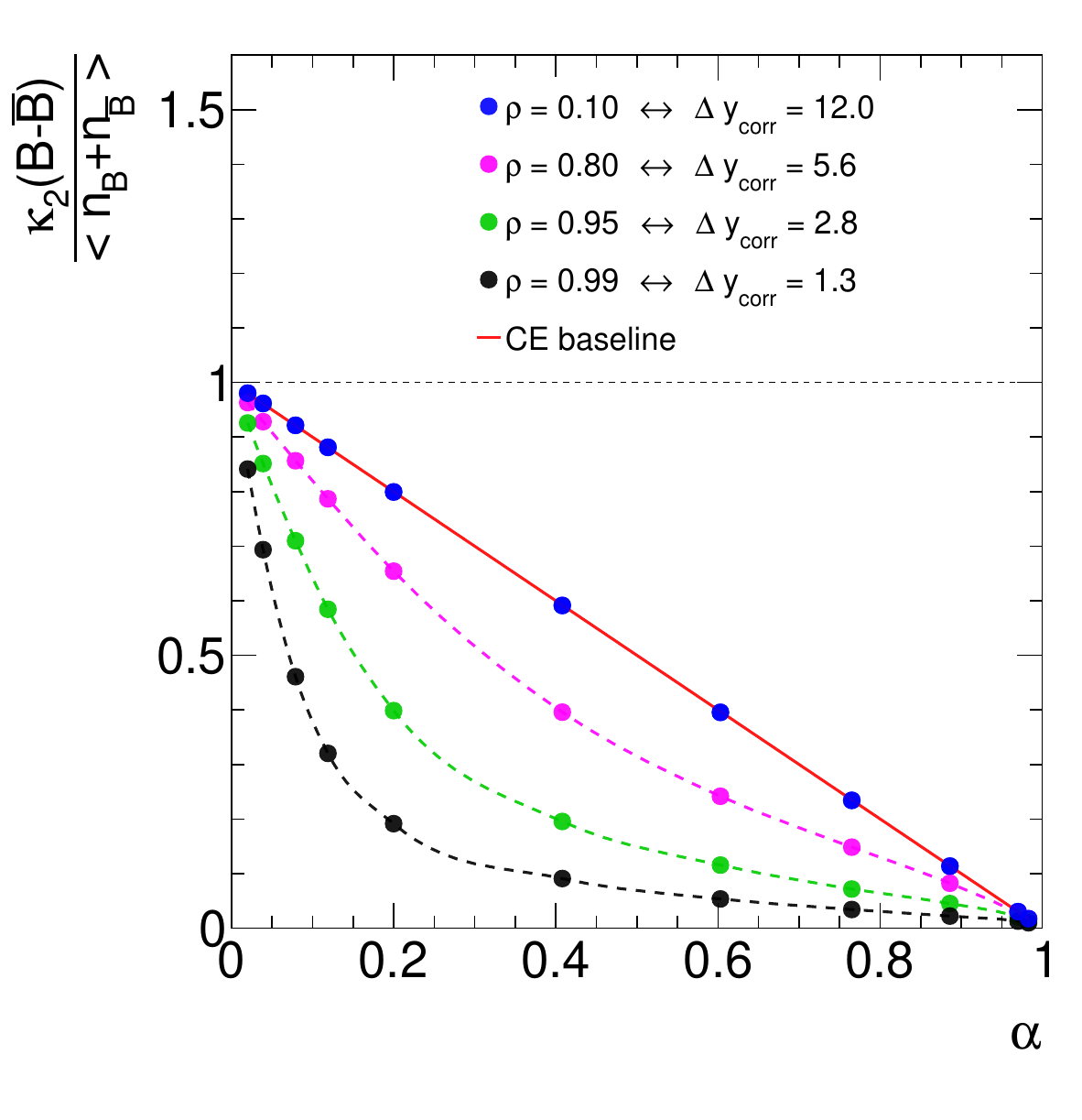}
 \includegraphics[width=0.45\linewidth,clip=true]{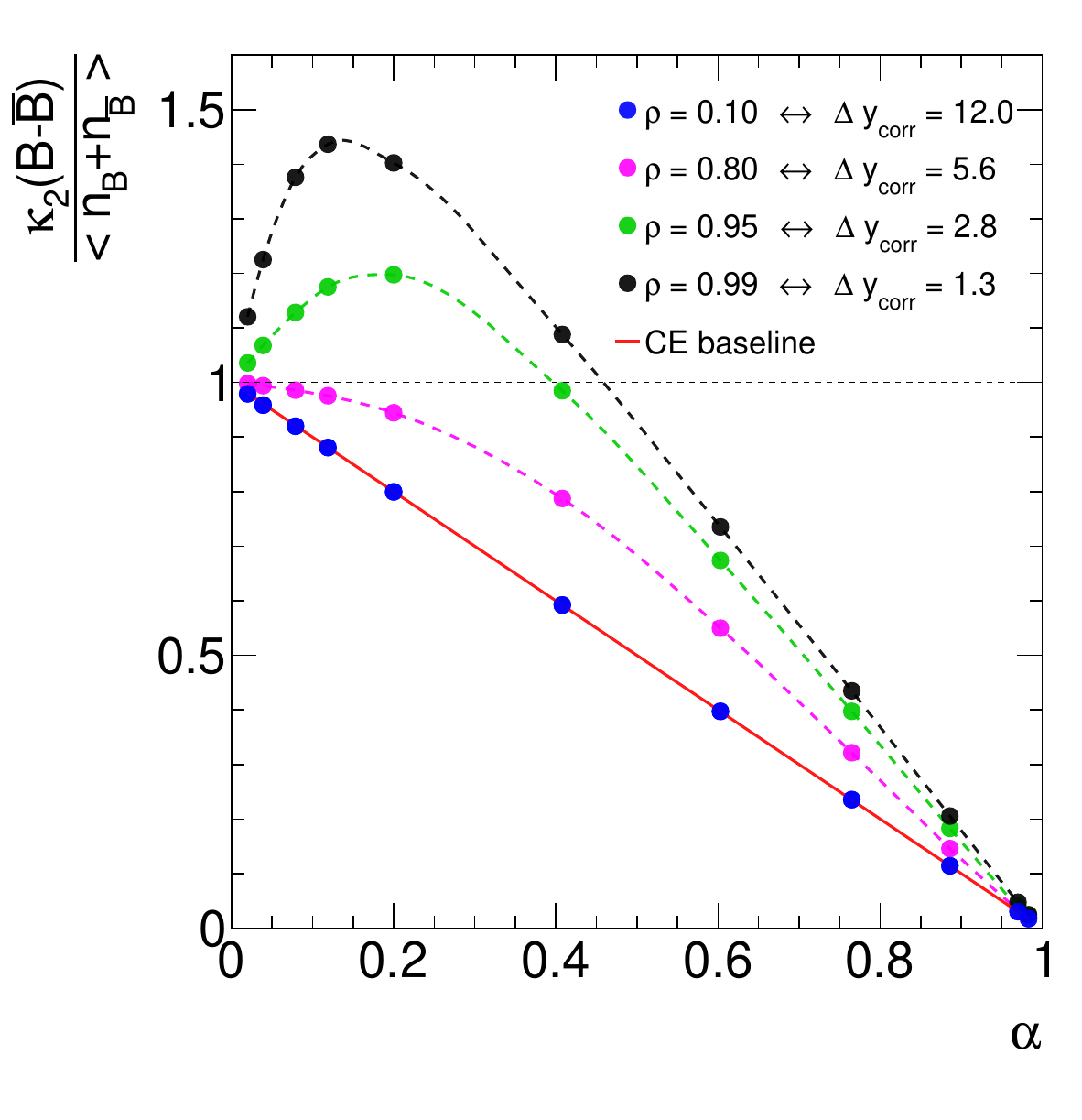}
 \caption{Second order net-baryon cumulants normalized to the mean number of baryons plus antibaryons as a function of the acceptance factor $\alpha$. The different color symbols correspond to different values of the correlation coefficient, or the correlation length in rapidity. In the left panel, correlations between unlike-sign charges (baryon-antibaryon) are introduced, while in the right panel the correlations are between like-sign charges (baryon-baryon or antibaryon-antibaryon). The solid red line corresponds to our previous calculation for global baryon number conservation ~\cite{Braun-Munzinger:2019yxj}.}
\label{fig:LHC}
\end{figure}

We start with the rapidity distributions of  charged particles as measured by the ALICE collaboration and further assume that in the central region, protons and antiprotons have equal rapidity distributions. Recent ALICE measurements of the antiproton/proton ratio in Pb--Pb collisions near mid-rapidity demonstrate that this ratio deviates from 1 by less than 1\%, independent of transverse momentum for central collisions ~\cite{ALICE:2023ulv}. Next, using the Metropolis algorithm, introduced in section~\ref{sec_metropolis} we induce correlations between baryons and anti-baryons in  rapidity space. Two examples with different correlation coefficients are presented in Fig.~\ref{ALICE04}, $\rho=0.4$ (left panel) and $\rho=0.8$ (right panel). The so obtained correlations between baryons and antibaryons in rapidity space with different correlation coefficients are then used in Eq.~\ref{eq:zCE} to generate baryon and antibaryon distributions as a function of experimental acceptance. Fig.~\ref{fig:LHC} represents the results obtained for normalized second order cumulants of net-baryons as a function of acceptance. The solid blue symbols correspond to a correlation coefficient $\rho = 0.1$ and are consistent with our previous calculations for global baryon number conservation \cite{Braun-Munzinger:2019yxj, ALICE:2019nbs},
shown by the red solid line. 
As seen from Fig.~\ref{fig:LHC}, in full acceptance fluctuations of net-baryons vanish, while for small acceptance they approach the small number Poisson limit. For intermediate values of the acceptance the results strongly depend on the correlation coefficient. Moreover, inspection of Fig.~\ref{fig:LHC} demonstrates two important outcomes. As explained above, the larger the correlation coefficient between unlike-sign pairs, the smaller the fluctuations of net-baryons. This is what one observes in the left panel of Fig.~\ref{fig:LHC}. On the other hand, inclusion of correlations between like-sign pairs increases the fluctuations (see the left panel of Fig.~\ref{fig:LHC}). This is because, in this case, the probability of producing clusters of baryons or antibaryons increases with increasing correlation coefficient which leads to stronger fluctuations of the net-baryon number. Naturally, a correlation between like-sign baryons should also be strongly manifested in the cumulants of the total baryon distribution. 

\section{Comparison to the ALICE data}
\begin{figure}[htb]
\centering
 \includegraphics[width=0.8\linewidth,clip=true]{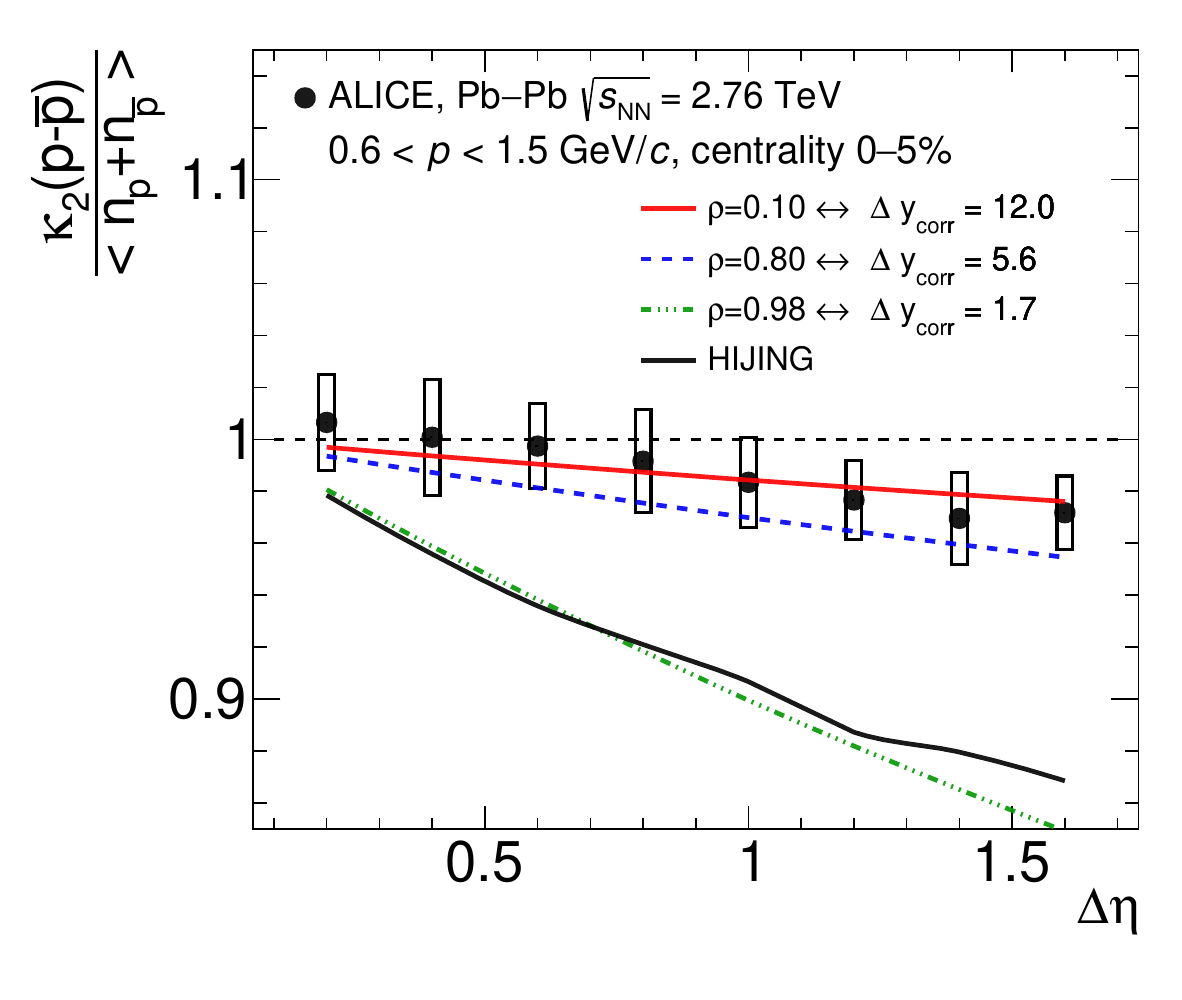}
 \caption{Second order cumulants of the net-proton distribution normalized to the mean number of protons plus antiprotons as a function of the pseudo-rapidity interval $\Delta\eta$ or acceptance fraction $\alpha(\Delta\eta)$, see text. The lines correspond to different values of the correlation coefficient between baryons and anti-baryons, or the correlation length in rapidity. For information on the HIJING calculation, see ~\cite{Gyulassy:1994ew}.}
\label{fig:ALICE-Data}
\end{figure}

The acceptance dependence of the normalized second order cumulants of net-proton number, as measured by the ALICE collaboration, is presented in Fig.~\ref{fig:ALICE-Data} by the solid black symbols~\cite{ALICE:2019nbs}. The acceptance is varied by changing the pseudorapidity interval $\Delta \eta$ around mid-rapidity. With increasing acceptance the data exhibit a progressive deviation from unity towards lower values, reaching a value of 0.97 at $\Delta \eta$ = 1.6. We note that, within the GCE in the Boltzmann limit with the HRG  EoS, the $\kappa_{2}(p-\bar{p})/\kappa_{1}(p+\bar{p})$ ratio equals unity, independent of the acceptance. The approach to unity for small values of $\Delta \eta$ is essentially driven by small number Poisson statistics.
In order to compare to the results from the previous section (see left panel of Fig.~\ref{fig:LHC}) we first estimated the acceptance fraction $\alpha$ corresponding to each experimental $\Delta \eta$, the $p_T$ range of the experimental data and the fraction of all baryons represented by the protons and antiprotons. The $\rho \approx 0$ case (numerically very close to $\rho = 0.1$ and not shown)
corresponds to missing correlations between baryon-antibaryon in rapidity, equivalent to global baryon number conservation. Larger values of $\rho$, as shown for $\rho$ = 0.80 and 0.98 would be indicative of local correlations of shorter range of 5.6 or 1.7 units of rapidity. The red solid line in Fig.~\ref{fig:ALICE-Data}, corresponding to our simulations with $\rho$ = 0.1 or $\Delta y_{corr}$ = 12, describes best the experimental data. 
The ALICE data imply long range correlations, consistent with our previous calculations~\cite{Braun-Munzinger:2019yxj, Braun-Munzinger:2018yru, Braun-Munzinger:2020jbk}. The shortest range correlation consistent with the experimental uncertainties within one standard deviation is 5.6 units in rapidity represented by the blue dashed line with $\rho$ = 0.8 shown in Fig.~\ref{fig:ALICE-Data}. For a tighter constraint, high precision measurements of second- and higher-order cumulants over a wider acceptance range are needed. In Fig.~\ref{fig:ALICE-Data} we also show correlations predicted using the HIJING event generator ~\cite{Gyulassy:1994ew}
(black solid line), which uses the Lund string model for baryon production~\cite{Andersson:1983ia}. The large deviation between the HIJING results and the experimental data calls into question the baryon production mechanism implemented via string fragmentation. This fragmentation mechanism would lead  to strong correlations in rapidity space~\cite{Eden:1996xi}, at variance with the ALICE data. Indeed, the HIJING results can be closely described by using a rather large correlation coefficient $\rho$ = 0.98. 
Comparison of the ALICE data  presented in Fig.~\ref{fig:ALICE-Data} with the right panel of Fig.~\ref{fig:LHC} does not support strong correlations between like-sign pairs. The situation can, however, change dramatically when considering lower energy collisions, where protons partially originate from fragmentation of the incoming nuclei. In this case, a distinction has to be made between stopped and produced protons while implementing local baryon number conservation. This is one of the unresolved issues in the field of particle number fluctuations. Nuclear collisions at intermediate energies would ideally be suitable for such studies. 

\section{Conclusions}
In summary, we have studied the effects of local baryon number conservation on fluctuations of the net-baryon and net-proton number at LHC energy. We first developed methods for introducing correlations in rapidity space between like- and unlike-sign baryon pairs. Generally, the method proposed here can be used to discriminate between different scenarios for baryon production and correlations in nuclear collisions. This will be particularly important for the analysis of data at lower energies, where such correlations might be very important.
Next, while imposing strict baryon number conservation in the overall volume, using the CE formulation of statistical mechanics, we studied baryon number fluctuations as a function of acceptance and the correlation coefficient or correlation length. In the absence of correlations, our approach is equivalent to independent binomial losses of baryons and anti-baryons as it was used in our previous studies~\cite{Braun-Munzinger:2019yxj, Braun-Munzinger:2018yru, Braun-Munzinger:2020jbk}.  By comparing our calculations with the measurements of the ALICE data we conclude that the correlation coefficient is below 0.8, corresponding to correlation lengths in rapidity larger than 5.6 units. The ALICE data lend support to the interpretation that the rapidity correlations between baryons and anti-baryons are long range correlations. On the other hand, the corresponding results from the HIJING event generator can be explained by introducing a correlation coefficient of $\rho$ = 0.98, corresponding to correlation lengths in rapidity smaller than 1.7 units. Such rather strong local correlations in HIJING are at odds with the ALICE data. This calls into question the baryon production mechanism implemented in string fragmentation models, such as the Lund string model. Further studies of fluctuations are needed to address also the possible role of baryon antibaryon annihilation ~\cite{Savchuk:2021aog,Vovchenko:2022xil} as well as time evolution of fluctuations in the hadronic medium~\cite{Kitazawa:2015ira, Sakaida:2014pya}.

\section*{Acknowledgments}
This work is part of and was supported by the DFG Collaborative Research
Centre "SFB 1225 (ISOQUANT)". K.R. acknowledges the support from the Polish National Science Center (NCN) under Opus grant no. 2022/45/B/ST2/01527  and of the Polish Ministry of Science and Higher Education.

\bibliography{LocalConservations}

\begin{thebibliography}{67}%
\makeatletter
\providecommand \@ifxundefined [1]{%
 \@ifx{#1\undefined}
}%
\providecommand \@ifnum [1]{%
 \ifnum #1\expandafter \@firstoftwo
 \else \expandafter \@secondoftwo
 \fi
}%
\providecommand \@ifx [1]{%
 \ifx #1\expandafter \@firstoftwo
 \else \expandafter \@secondoftwo
 \fi
}%
\providecommand \natexlab [1]{#1}%
\providecommand \enquote  [1]{``#1''}%
\providecommand \bibnamefont  [1]{#1}%
\providecommand \bibfnamefont [1]{#1}%
\providecommand \citenamefont [1]{#1}%
\providecommand \href@noop [0]{\@secondoftwo}%
\providecommand \href [0]{\begingroup \@sanitize@url \@href}%
\providecommand \@href[1]{\@@startlink{#1}\@@href}%
\providecommand \@@href[1]{\endgroup#1\@@endlink}%
\providecommand \@sanitize@url [0]{\catcode `\\12\catcode `\$12\catcode
  `\&12\catcode `\#12\catcode `\^12\catcode `\_12\catcode `\%12\relax}%
\providecommand \@@startlink[1]{}%
\providecommand \@@endlink[0]{}%
\providecommand \url  [0]{\begingroup\@sanitize@url \@url }%
\providecommand \@url [1]{\endgroup\@href {#1}{\urlprefix }}%
\providecommand \urlprefix  [0]{URL }%
\providecommand \Eprint [0]{\href }%
\providecommand \doibase [0]{http://dx.doi.org/}%
\providecommand \selectlanguage [0]{\@gobble}%
\providecommand \bibinfo  [0]{\@secondoftwo}%
\providecommand \bibfield  [0]{\@secondoftwo}%
\providecommand \translation [1]{[#1]}%
\providecommand \BibitemOpen [0]{}%
\providecommand \bibitemStop [0]{}%
\providecommand \bibitemNoStop [0]{.\EOS\space}%
\providecommand \EOS [0]{\spacefactor3000\relax}%
\providecommand \BibitemShut  [1]{\csname bibitem#1\endcsname}%
\let\auto@bib@innerbib\@empty
\bibitem [{\citenamefont {Braun-Munzinger}\ \emph {et~al.}(2022)\citenamefont
  {Braun-Munzinger}, \citenamefont {Rustamov},\ and\ \citenamefont
  {Stachel}}]{Braun-Munzinger:2022bkc}%
  \BibitemOpen
  \bibfield  {author} {\bibinfo {author} {\bibfnamefont {P.}~\bibnamefont
  {Braun-Munzinger}}, \bibinfo {author} {\bibfnamefont {A.}~\bibnamefont
  {Rustamov}}, \ and\ \bibinfo {author} {\bibfnamefont {J.}~\bibnamefont
  {Stachel}},\ }\href@noop {} {\  (\bibinfo {year} {2022})},\ \Eprint
  {http://arxiv.org/abs/2211.08819} {arXiv:2211.08819 [hep-ph]} \BibitemShut
  {NoStop}%
\bibitem [{\citenamefont {Gross}\ \emph {et~al.}(2022)\citenamefont {Gross}
  \emph {et~al.}}]{Gross:2022hyw}%
  \BibitemOpen
  \bibfield  {author} {\bibinfo {author} {\bibfnamefont {F.}~\bibnamefont
  {Gross}} \emph {et~al.},\ }\href@noop {} {\  (\bibinfo {year} {2022})},\
  \Eprint {http://arxiv.org/abs/2212.11107} {arXiv:2212.11107 [hep-ph]}
  \BibitemShut {NoStop}%
\bibitem [{\citenamefont {Andronic}\ \emph {et~al.}(2018)\citenamefont
  {Andronic}, \citenamefont {Braun-Munzinger}, \citenamefont {Redlich},\ and\
  \citenamefont {Stachel}}]{HRG}%
  \BibitemOpen
  \bibfield  {author} {\bibinfo {author} {\bibfnamefont {A.}~\bibnamefont
  {Andronic}}, \bibinfo {author} {\bibfnamefont {P.}~\bibnamefont
  {Braun-Munzinger}}, \bibinfo {author} {\bibfnamefont {K.}~\bibnamefont
  {Redlich}}, \ and\ \bibinfo {author} {\bibfnamefont {J.}~\bibnamefont
  {Stachel}},\ }\href {\doibase 10.1038/s41586-018-0491-6} {\bibfield
  {journal} {\bibinfo  {journal} {Nature}\ }\textbf {\bibinfo {volume} {561}},\
  \bibinfo {pages} {321} (\bibinfo {year} {2018})},\ \Eprint
  {http://arxiv.org/abs/1710.09425} {arXiv:1710.09425 [nucl-th]} \BibitemShut
  {NoStop}%
\bibitem [{\citenamefont {Andronic}\ \emph {et~al.}(2006)\citenamefont
  {Andronic}, \citenamefont {Braun-Munzinger},\ and\ \citenamefont
  {Stachel}}]{Andronic:2005yp}%
  \BibitemOpen
  \bibfield  {author} {\bibinfo {author} {\bibfnamefont {A.}~\bibnamefont
  {Andronic}}, \bibinfo {author} {\bibfnamefont {P.}~\bibnamefont
  {Braun-Munzinger}}, \ and\ \bibinfo {author} {\bibfnamefont {J.}~\bibnamefont
  {Stachel}},\ }\href {\doibase 10.1016/j.nuclphysa.2006.03.012} {\bibfield
  {journal} {\bibinfo  {journal} {Nucl. Phys. A}\ }\textbf {\bibinfo {volume}
  {772}},\ \bibinfo {pages} {167} (\bibinfo {year} {2006})},\ \Eprint
  {http://arxiv.org/abs/nucl-th/0511071} {arXiv:nucl-th/0511071} \BibitemShut
  {NoStop}%
\bibitem [{\citenamefont {Andronic}\ \emph {et~al.}(2019)\citenamefont
  {Andronic}, \citenamefont {Braun-Munzinger}, \citenamefont {Friman},
  \citenamefont {Lo}, \citenamefont {Redlich},\ and\ \citenamefont
  {Stachel}}]{Andronic:2018qqt}%
  \BibitemOpen
  \bibfield  {author} {\bibinfo {author} {\bibfnamefont {A.}~\bibnamefont
  {Andronic}}, \bibinfo {author} {\bibfnamefont {P.}~\bibnamefont
  {Braun-Munzinger}}, \bibinfo {author} {\bibfnamefont {B.}~\bibnamefont
  {Friman}}, \bibinfo {author} {\bibfnamefont {P.~M.}\ \bibnamefont {Lo}},
  \bibinfo {author} {\bibfnamefont {K.}~\bibnamefont {Redlich}}, \ and\
  \bibinfo {author} {\bibfnamefont {J.}~\bibnamefont {Stachel}},\ }\href
  {\doibase 10.1016/j.physletb.2019.03.052} {\bibfield  {journal} {\bibinfo
  {journal} {Phys. Lett.}\ }\textbf {\bibinfo {volume} {B792}},\ \bibinfo
  {pages} {304} (\bibinfo {year} {2019})},\ \Eprint
  {http://arxiv.org/abs/1808.03102} {arXiv:1808.03102 [hep-ph]} \BibitemShut
  {NoStop}%
\bibitem [{\citenamefont {Asakawa}\ and\ \citenamefont
  {Yazaki}(1989)}]{Asakawa:1989bq}%
  \BibitemOpen
  \bibfield  {author} {\bibinfo {author} {\bibfnamefont {M.}~\bibnamefont
  {Asakawa}}\ and\ \bibinfo {author} {\bibfnamefont {K.}~\bibnamefont
  {Yazaki}},\ }\href {\doibase 10.1016/0375-9474(89)90002-X} {\bibfield
  {journal} {\bibinfo  {journal} {Nucl. Phys. A}\ }\textbf {\bibinfo {volume}
  {504}},\ \bibinfo {pages} {668} (\bibinfo {year} {1989})}\BibitemShut
  {NoStop}%
\bibitem [{\citenamefont {Rajagopal}\ and\ \citenamefont
  {Wilczek}(1993)}]{Rajagopal:1992qz}%
  \BibitemOpen
  \bibfield  {author} {\bibinfo {author} {\bibfnamefont {K.}~\bibnamefont
  {Rajagopal}}\ and\ \bibinfo {author} {\bibfnamefont {F.}~\bibnamefont
  {Wilczek}},\ }\href {\doibase 10.1016/0550-3213(93)90502-G} {\bibfield
  {journal} {\bibinfo  {journal} {Nucl. Phys. B}\ }\textbf {\bibinfo {volume}
  {399}},\ \bibinfo {pages} {395} (\bibinfo {year} {1993})},\ \Eprint
  {http://arxiv.org/abs/hep-ph/9210253} {arXiv:hep-ph/9210253} \BibitemShut
  {NoStop}%
\bibitem [{\citenamefont {Stephanov}\ \emph {et~al.}(1998)\citenamefont
  {Stephanov}, \citenamefont {Rajagopal},\ and\ \citenamefont
  {Shuryak}}]{Stephanov:1998dy}%
  \BibitemOpen
  \bibfield  {author} {\bibinfo {author} {\bibfnamefont {M.~A.}\ \bibnamefont
  {Stephanov}}, \bibinfo {author} {\bibfnamefont {K.}~\bibnamefont
  {Rajagopal}}, \ and\ \bibinfo {author} {\bibfnamefont {E.~V.}\ \bibnamefont
  {Shuryak}},\ }\href {\doibase 10.1103/PhysRevLett.81.4816} {\bibfield
  {journal} {\bibinfo  {journal} {Phys. Rev. Lett.}\ }\textbf {\bibinfo
  {volume} {81}},\ \bibinfo {pages} {4816} (\bibinfo {year} {1998})},\ \Eprint
  {http://arxiv.org/abs/hep-ph/9806219} {arXiv:hep-ph/9806219} \BibitemShut
  {NoStop}%
\bibitem [{\citenamefont {Fu}\ \emph {et~al.}(2020)\citenamefont {Fu},
  \citenamefont {Pawlowski},\ and\ \citenamefont {Rennecke}}]{Fu:2019hdw}%
  \BibitemOpen
  \bibfield  {author} {\bibinfo {author} {\bibfnamefont {W.-j.}\ \bibnamefont
  {Fu}}, \bibinfo {author} {\bibfnamefont {J.~M.}\ \bibnamefont {Pawlowski}}, \
  and\ \bibinfo {author} {\bibfnamefont {F.}~\bibnamefont {Rennecke}},\ }\href
  {\doibase 10.1103/PhysRevD.101.054032} {\bibfield  {journal} {\bibinfo
  {journal} {Phys. Rev. D}\ }\textbf {\bibinfo {volume} {101}},\ \bibinfo
  {pages} {054032} (\bibinfo {year} {2020})},\ \Eprint
  {http://arxiv.org/abs/1909.02991} {arXiv:1909.02991 [hep-ph]} \BibitemShut
  {NoStop}%
\bibitem [{\citenamefont {Bazavov}\ \emph {et~al.}(2012)\citenamefont {Bazavov}
  \emph {et~al.}}]{LQCD1}%
  \BibitemOpen
  \bibfield  {author} {\bibinfo {author} {\bibfnamefont {A.}~\bibnamefont
  {Bazavov}} \emph {et~al.},\ }\href {\doibase 10.1103/PhysRevD.85.054503}
  {\bibfield  {journal} {\bibinfo  {journal} {Phys. Rev.}\ }\textbf {\bibinfo
  {volume} {D85}},\ \bibinfo {pages} {054503} (\bibinfo {year} {2012})},\
  \Eprint {http://arxiv.org/abs/1111.1710} {arXiv:1111.1710 [hep-lat]}
  \BibitemShut {NoStop}%
\bibitem [{\citenamefont {Friman}\ \emph {et~al.}(2011)\citenamefont {Friman},
  \citenamefont {Karsch}, \citenamefont {Redlich},\ and\ \citenamefont
  {Skokov}}]{Redlich1}%
  \BibitemOpen
  \bibfield  {author} {\bibinfo {author} {\bibfnamefont {B.}~\bibnamefont
  {Friman}}, \bibinfo {author} {\bibfnamefont {F.}~\bibnamefont {Karsch}},
  \bibinfo {author} {\bibfnamefont {K.}~\bibnamefont {Redlich}}, \ and\
  \bibinfo {author} {\bibfnamefont {V.}~\bibnamefont {Skokov}},\ }\href
  {\doibase 10.1140/epjc/s10052-011-1694-2} {\bibfield  {journal} {\bibinfo
  {journal} {Eur. Phys. J.}\ }\textbf {\bibinfo {volume} {C71}},\ \bibinfo
  {pages} {1694} (\bibinfo {year} {2011})},\ \Eprint
  {http://arxiv.org/abs/1103.3511} {arXiv:1103.3511 [hep-ph]} \BibitemShut
  {NoStop}%
\bibitem [{\citenamefont {Bazavov}\ \emph {et~al.}(2019)\citenamefont {Bazavov}
  \emph {et~al.}}]{HotQCD:2018pds}%
  \BibitemOpen
  \bibfield  {author} {\bibinfo {author} {\bibfnamefont {A.}~\bibnamefont
  {Bazavov}} \emph {et~al.} (\bibinfo {collaboration} {HotQCD}),\ }\href
  {\doibase 10.1016/j.physletb.2019.05.013} {\bibfield  {journal} {\bibinfo
  {journal} {Phys. Lett. B}\ }\textbf {\bibinfo {volume} {795}},\ \bibinfo
  {pages} {15} (\bibinfo {year} {2019})},\ \Eprint
  {http://arxiv.org/abs/1812.08235} {arXiv:1812.08235 [hep-lat]} \BibitemShut
  {NoStop}%
\bibitem [{\citenamefont {Borsanyi}\ \emph {et~al.}(2020)\citenamefont
  {Borsanyi}, \citenamefont {Fodor}, \citenamefont {Guenther}, \citenamefont
  {Kara}, \citenamefont {Katz}, \citenamefont {Parotto}, \citenamefont
  {Pasztor}, \citenamefont {Ratti},\ and\ \citenamefont
  {Szabo}}]{Borsanyi:2020fev}%
  \BibitemOpen
  \bibfield  {author} {\bibinfo {author} {\bibfnamefont {S.}~\bibnamefont
  {Borsanyi}}, \bibinfo {author} {\bibfnamefont {Z.}~\bibnamefont {Fodor}},
  \bibinfo {author} {\bibfnamefont {J.~N.}\ \bibnamefont {Guenther}}, \bibinfo
  {author} {\bibfnamefont {R.}~\bibnamefont {Kara}}, \bibinfo {author}
  {\bibfnamefont {S.~D.}\ \bibnamefont {Katz}}, \bibinfo {author}
  {\bibfnamefont {P.}~\bibnamefont {Parotto}}, \bibinfo {author} {\bibfnamefont
  {A.}~\bibnamefont {Pasztor}}, \bibinfo {author} {\bibfnamefont
  {C.}~\bibnamefont {Ratti}}, \ and\ \bibinfo {author} {\bibfnamefont {K.~K.}\
  \bibnamefont {Szabo}},\ }\href {\doibase 10.1103/PhysRevLett.125.052001}
  {\bibfield  {journal} {\bibinfo  {journal} {Phys. Rev. Lett.}\ }\textbf
  {\bibinfo {volume} {125}},\ \bibinfo {pages} {052001} (\bibinfo {year}
  {2020})},\ \Eprint {http://arxiv.org/abs/2002.02821} {arXiv:2002.02821
  [hep-lat]} \BibitemShut {NoStop}%
\bibitem [{\citenamefont {Pisarski}\ and\ \citenamefont
  {Wilczek}(1984)}]{Pisarski:1983ms}%
  \BibitemOpen
  \bibfield  {author} {\bibinfo {author} {\bibfnamefont {R.~D.}\ \bibnamefont
  {Pisarski}}\ and\ \bibinfo {author} {\bibfnamefont {F.}~\bibnamefont
  {Wilczek}},\ }\href {\doibase 10.1103/PhysRevD.29.338} {\bibfield  {journal}
  {\bibinfo  {journal} {Phys. Rev. D}\ }\textbf {\bibinfo {volume} {29}},\
  \bibinfo {pages} {338} (\bibinfo {year} {1984})}\BibitemShut {NoStop}%
\bibitem [{\citenamefont {Ejiri}\ \emph {et~al.}(2006)\citenamefont {Ejiri},
  \citenamefont {Karsch},\ and\ \citenamefont {Redlich}}]{Ejiri:2005wq}%
  \BibitemOpen
  \bibfield  {author} {\bibinfo {author} {\bibfnamefont {S.}~\bibnamefont
  {Ejiri}}, \bibinfo {author} {\bibfnamefont {F.}~\bibnamefont {Karsch}}, \
  and\ \bibinfo {author} {\bibfnamefont {K.}~\bibnamefont {Redlich}},\ }\href
  {\doibase 10.1016/j.physletb.2005.11.083} {\bibfield  {journal} {\bibinfo
  {journal} {Phys. Lett. B}\ }\textbf {\bibinfo {volume} {633}},\ \bibinfo
  {pages} {275} (\bibinfo {year} {2006})},\ \Eprint
  {http://arxiv.org/abs/hep-ph/0509051} {arXiv:hep-ph/0509051} \BibitemShut
  {NoStop}%
\bibitem [{\citenamefont {Adamczewski-Musch}\ \emph {et~al.}(2020)\citenamefont
  {Adamczewski-Musch} \emph {et~al.}}]{HADES:2020wpc}%
  \BibitemOpen
  \bibfield  {author} {\bibinfo {author} {\bibfnamefont {J.}~\bibnamefont
  {Adamczewski-Musch}} \emph {et~al.} (\bibinfo {collaboration} {HADES}),\
  }\href {\doibase 10.1103/PhysRevC.102.024914} {\bibfield  {journal} {\bibinfo
   {journal} {Phys. Rev. C}\ }\textbf {\bibinfo {volume} {102}},\ \bibinfo
  {pages} {024914} (\bibinfo {year} {2020})},\ \Eprint
  {http://arxiv.org/abs/2002.08701} {arXiv:2002.08701 [nucl-ex]} \BibitemShut
  {NoStop}%
\bibitem [{\citenamefont {Adam}\ \emph {et~al.}(2021)\citenamefont {Adam} \emph
  {et~al.}}]{STAR:2020tga}%
  \BibitemOpen
  \bibfield  {author} {\bibinfo {author} {\bibfnamefont {J.}~\bibnamefont
  {Adam}} \emph {et~al.} (\bibinfo {collaboration} {STAR}),\ }\href {\doibase
  10.1103/PhysRevLett.126.092301} {\bibfield  {journal} {\bibinfo  {journal}
  {Phys. Rev. Lett.}\ }\textbf {\bibinfo {volume} {126}},\ \bibinfo {pages}
  {092301} (\bibinfo {year} {2021})},\ \Eprint
  {http://arxiv.org/abs/2001.02852} {arXiv:2001.02852 [nucl-ex]} \BibitemShut
  {NoStop}%
\bibitem [{\citenamefont {Abdallah}\ \emph {et~al.}(2022)\citenamefont
  {Abdallah} \emph {et~al.}}]{STAR:2021fge}%
  \BibitemOpen
  \bibfield  {author} {\bibinfo {author} {\bibfnamefont {M.~S.}\ \bibnamefont
  {Abdallah}} \emph {et~al.} (\bibinfo {collaboration} {STAR}),\ }\href
  {\doibase 10.1103/PhysRevLett.128.202303} {\bibfield  {journal} {\bibinfo
  {journal} {Phys. Rev. Lett.}\ }\textbf {\bibinfo {volume} {128}},\ \bibinfo
  {pages} {202303} (\bibinfo {year} {2022})},\ \Eprint
  {http://arxiv.org/abs/2112.00240} {arXiv:2112.00240 [nucl-ex]} \BibitemShut
  {NoStop}%
\bibitem [{\citenamefont {Abdallah}\ \emph
  {et~al.}(2021{\natexlab{a}})\citenamefont {Abdallah} \emph
  {et~al.}}]{STAR:2021rls}%
  \BibitemOpen
  \bibfield  {author} {\bibinfo {author} {\bibfnamefont {M.}~\bibnamefont
  {Abdallah}} \emph {et~al.} (\bibinfo {collaboration} {STAR}),\ }\href
  {\doibase 10.1103/PhysRevLett.127.262301} {\bibfield  {journal} {\bibinfo
  {journal} {Phys. Rev. Lett.}\ }\textbf {\bibinfo {volume} {127}},\ \bibinfo
  {pages} {262301} (\bibinfo {year} {2021}{\natexlab{a}})},\ \Eprint
  {http://arxiv.org/abs/2105.14698} {arXiv:2105.14698 [nucl-ex]} \BibitemShut
  {NoStop}%
\bibitem [{\citenamefont {Abdallah}\ \emph
  {et~al.}(2021{\natexlab{b}})\citenamefont {Abdallah} \emph
  {et~al.}}]{STAR:2021iop}%
  \BibitemOpen
  \bibfield  {author} {\bibinfo {author} {\bibfnamefont {M.}~\bibnamefont
  {Abdallah}} \emph {et~al.} (\bibinfo {collaboration} {STAR}),\ }\href
  {\doibase 10.1103/PhysRevC.104.024902} {\bibfield  {journal} {\bibinfo
  {journal} {Phys. Rev. C}\ }\textbf {\bibinfo {volume} {104}},\ \bibinfo
  {pages} {024902} (\bibinfo {year} {2021}{\natexlab{b}})},\ \Eprint
  {http://arxiv.org/abs/2101.12413} {arXiv:2101.12413 [nucl-ex]} \BibitemShut
  {NoStop}%
\bibitem [{\citenamefont {Xu}(2018)}]{Xu:2018vnf}%
  \BibitemOpen
  \bibfield  {author} {\bibinfo {author} {\bibfnamefont {N.}~\bibnamefont
  {Xu}},\ }\bibfield  {booktitle} {\emph {\bibinfo {booktitle} {{Proceedings,
  XXII DAE High Energy Physics Symposium: Delhi, India, December 12 -16,
  2016}}},\ }\href {\doibase 10.1007/978-3-319-73171-1_1} {\bibfield  {journal}
  {\bibinfo  {journal} {Springer Proc. Phys.}\ }\textbf {\bibinfo {volume}
  {203}},\ \bibinfo {pages} {1} (\bibinfo {year} {2018})}\BibitemShut {NoStop}%
\bibitem [{\citenamefont {Abdallah}\ \emph {et~al.}(2023)\citenamefont
  {Abdallah} \emph {et~al.}}]{STAR:2022etb}%
  \BibitemOpen
  \bibfield  {author} {\bibinfo {author} {\bibfnamefont {M.}~\bibnamefont
  {Abdallah}} \emph {et~al.} (\bibinfo {collaboration} {STAR}),\ }\href
  {\doibase 10.1103/PhysRevC.107.024908} {\bibfield  {journal} {\bibinfo
  {journal} {Phys. Rev. C}\ }\textbf {\bibinfo {volume} {107}},\ \bibinfo
  {pages} {024908} (\bibinfo {year} {2023})},\ \Eprint
  {http://arxiv.org/abs/2209.11940} {arXiv:2209.11940 [nucl-ex]} \BibitemShut
  {NoStop}%
\bibitem [{\citenamefont {Rustamov}(2017)}]{RustamovQM17}%
  \BibitemOpen
  \bibfield  {author} {\bibinfo {author} {\bibfnamefont {A.}~\bibnamefont
  {Rustamov}} (\bibinfo {collaboration} {ALICE}),\ }\bibfield  {booktitle}
  {\emph {\bibinfo {booktitle} {{Proceedings, 26th International Conference on
  Ultra-relativistic Nucleus-Nucleus Collisions (Quark Matter 2017): Chicago,
  Illinois, USA, February 5-11, 2017}}},\ }\href {\doibase
  10.1016/j.nuclphysa.2017.05.111} {\bibfield  {journal} {\bibinfo  {journal}
  {Nucl. Phys.}\ }\textbf {\bibinfo {volume} {A967}},\ \bibinfo {pages} {453}
  (\bibinfo {year} {2017})},\ \Eprint {http://arxiv.org/abs/1704.05329}
  {arXiv:1704.05329 [nucl-ex]} \BibitemShut {NoStop}%
\bibitem [{\citenamefont {Acharya}\ \emph {et~al.}(2020)\citenamefont {Acharya}
  \emph {et~al.}}]{ALICE:2019nbs}%
  \BibitemOpen
  \bibfield  {author} {\bibinfo {author} {\bibfnamefont {S.}~\bibnamefont
  {Acharya}} \emph {et~al.} (\bibinfo {collaboration} {ALICE}),\ }\href
  {\doibase 10.1016/j.physletb.2020.135564} {\bibfield  {journal} {\bibinfo
  {journal} {Phys. Lett. B}\ }\textbf {\bibinfo {volume} {807}},\ \bibinfo
  {pages} {135564} (\bibinfo {year} {2020})},\ \Eprint
  {http://arxiv.org/abs/1910.14396} {arXiv:1910.14396 [nucl-ex]} \BibitemShut
  {NoStop}%
\bibitem [{\citenamefont {Acharya}\ \emph
  {et~al.}(2023{\natexlab{a}})\citenamefont {Acharya} \emph
  {et~al.}}]{ALICE:2022xpf}%
  \BibitemOpen
  \bibfield  {author} {\bibinfo {author} {\bibfnamefont {S.}~\bibnamefont
  {Acharya}} \emph {et~al.} (\bibinfo {collaboration} {ALICE}),\ }\href
  {\doibase 10.1016/j.physletb.2022.137545} {\bibfield  {journal} {\bibinfo
  {journal} {Phys. Lett. B}\ }\textbf {\bibinfo {volume} {844}},\ \bibinfo
  {pages} {137545} (\bibinfo {year} {2023}{\natexlab{a}})},\ \Eprint
  {http://arxiv.org/abs/2206.03343} {arXiv:2206.03343 [nucl-ex]} \BibitemShut
  {NoStop}%
\bibitem [{\citenamefont {Gorenstein}\ and\ \citenamefont
  {Gazdzicki}(2011)}]{Gorenstein:2011vq}%
  \BibitemOpen
  \bibfield  {author} {\bibinfo {author} {\bibfnamefont {M.~I.}\ \bibnamefont
  {Gorenstein}}\ and\ \bibinfo {author} {\bibfnamefont {M.}~\bibnamefont
  {Gazdzicki}},\ }\href {\doibase 10.1103/PhysRevC.84.014904} {\bibfield
  {journal} {\bibinfo  {journal} {Phys. Rev. C}\ }\textbf {\bibinfo {volume}
  {84}},\ \bibinfo {pages} {014904} (\bibinfo {year} {2011})},\ \Eprint
  {http://arxiv.org/abs/1101.4865} {arXiv:1101.4865 [nucl-th]} \BibitemShut
  {NoStop}%
\bibitem [{\citenamefont {Skokov}\ \emph {et~al.}(2013)\citenamefont {Skokov},
  \citenamefont {Friman},\ and\ \citenamefont {Redlich}}]{Skokov:2012ds}%
  \BibitemOpen
  \bibfield  {author} {\bibinfo {author} {\bibfnamefont {V.}~\bibnamefont
  {Skokov}}, \bibinfo {author} {\bibfnamefont {B.}~\bibnamefont {Friman}}, \
  and\ \bibinfo {author} {\bibfnamefont {K.}~\bibnamefont {Redlich}},\ }\href
  {\doibase 10.1103/PhysRevC.88.034911} {\bibfield  {journal} {\bibinfo
  {journal} {Phys. Rev. C}\ }\textbf {\bibinfo {volume} {88}},\ \bibinfo
  {pages} {034911} (\bibinfo {year} {2013})},\ \Eprint
  {http://arxiv.org/abs/1205.4756} {arXiv:1205.4756 [hep-ph]} \BibitemShut
  {NoStop}%
\bibitem [{\citenamefont {Braun-Munzinger}\ \emph {et~al.}(2017)\citenamefont
  {Braun-Munzinger}, \citenamefont {Rustamov},\ and\ \citenamefont
  {Stachel}}]{Braun-Munzinger:2016yjz}%
  \BibitemOpen
  \bibfield  {author} {\bibinfo {author} {\bibfnamefont {P.}~\bibnamefont
  {Braun-Munzinger}}, \bibinfo {author} {\bibfnamefont {A.}~\bibnamefont
  {Rustamov}}, \ and\ \bibinfo {author} {\bibfnamefont {J.}~\bibnamefont
  {Stachel}},\ }\href {\doibase 10.1016/j.nuclphysa.2017.01.011} {\bibfield
  {journal} {\bibinfo  {journal} {Nucl. Phys. A}\ }\textbf {\bibinfo {volume}
  {960}},\ \bibinfo {pages} {114} (\bibinfo {year} {2017})},\ \Eprint
  {http://arxiv.org/abs/1612.00702} {arXiv:1612.00702 [nucl-th]} \BibitemShut
  {NoStop}%
\bibitem [{\citenamefont {Rustamov}\ \emph {et~al.}(2023)\citenamefont
  {Rustamov}, \citenamefont {Stroth},\ and\ \citenamefont
  {Holzmann}}]{Rustamov:2022sqm}%
  \BibitemOpen
  \bibfield  {author} {\bibinfo {author} {\bibfnamefont {A.}~\bibnamefont
  {Rustamov}}, \bibinfo {author} {\bibfnamefont {J.}~\bibnamefont {Stroth}}, \
  and\ \bibinfo {author} {\bibfnamefont {R.}~\bibnamefont {Holzmann}},\ }\href
  {\doibase 10.1016/j.nuclphysa.2023.122641} {\bibfield  {journal} {\bibinfo
  {journal} {Nucl. Phys. A}\ }\textbf {\bibinfo {volume} {1034}},\ \bibinfo
  {pages} {122641} (\bibinfo {year} {2023})},\ \Eprint
  {http://arxiv.org/abs/2211.14849} {arXiv:2211.14849 [nucl-th]} \BibitemShut
  {NoStop}%
\bibitem [{\citenamefont {Bzdak}\ \emph {et~al.}(2013)\citenamefont {Bzdak},
  \citenamefont {Koch},\ and\ \citenamefont {Skokov}}]{Bzdak:2012an}%
  \BibitemOpen
  \bibfield  {author} {\bibinfo {author} {\bibfnamefont {A.}~\bibnamefont
  {Bzdak}}, \bibinfo {author} {\bibfnamefont {V.}~\bibnamefont {Koch}}, \ and\
  \bibinfo {author} {\bibfnamefont {V.}~\bibnamefont {Skokov}},\ }\href
  {\doibase 10.1103/PhysRevC.87.014901} {\bibfield  {journal} {\bibinfo
  {journal} {Phys. Rev. C}\ }\textbf {\bibinfo {volume} {87}},\ \bibinfo
  {pages} {014901} (\bibinfo {year} {2013})},\ \Eprint
  {http://arxiv.org/abs/1203.4529} {arXiv:1203.4529 [hep-ph]} \BibitemShut
  {NoStop}%
\bibitem [{\citenamefont {Braun-Munzinger}\ \emph
  {et~al.}(2019{\natexlab{a}})\citenamefont {Braun-Munzinger}, \citenamefont
  {Rustamov},\ and\ \citenamefont {Stachel}}]{Braun-Munzinger:2018yru}%
  \BibitemOpen
  \bibfield  {author} {\bibinfo {author} {\bibfnamefont {P.}~\bibnamefont
  {Braun-Munzinger}}, \bibinfo {author} {\bibfnamefont {A.}~\bibnamefont
  {Rustamov}}, \ and\ \bibinfo {author} {\bibfnamefont {J.}~\bibnamefont
  {Stachel}},\ }\bibfield  {booktitle} {\emph {\bibinfo {booktitle}
  {{Proceedings, 27th International Conference on Ultrarelativistic
  Nucleus-Nucleus Collisions (Quark Matter 2018): Venice, Italy, May 14-19,
  2018}}},\ }\href {\doibase 10.1016/j.nuclphysa.2018.09.074} {\bibfield
  {journal} {\bibinfo  {journal} {Nucl. Phys.}\ }\textbf {\bibinfo {volume}
  {A982}},\ \bibinfo {pages} {307} (\bibinfo {year} {2019}{\natexlab{a}})},\
  \Eprint {http://arxiv.org/abs/1807.08927} {arXiv:1807.08927 [nucl-th]}
  \BibitemShut {NoStop}%
\bibitem [{\citenamefont {Braun-Munzinger}\ \emph {et~al.}(2021)\citenamefont
  {Braun-Munzinger}, \citenamefont {Friman}, \citenamefont {Redlich},
  \citenamefont {Rustamov},\ and\ \citenamefont
  {Stachel}}]{Braun-Munzinger:2020jbk}%
  \BibitemOpen
  \bibfield  {author} {\bibinfo {author} {\bibfnamefont {P.}~\bibnamefont
  {Braun-Munzinger}}, \bibinfo {author} {\bibfnamefont {B.}~\bibnamefont
  {Friman}}, \bibinfo {author} {\bibfnamefont {K.}~\bibnamefont {Redlich}},
  \bibinfo {author} {\bibfnamefont {A.}~\bibnamefont {Rustamov}}, \ and\
  \bibinfo {author} {\bibfnamefont {J.}~\bibnamefont {Stachel}},\ }\href
  {\doibase 10.1016/j.nuclphysa.2021.122141} {\bibfield  {journal} {\bibinfo
  {journal} {Nucl. Phys. A}\ }\textbf {\bibinfo {volume} {1008}},\ \bibinfo
  {pages} {122141} (\bibinfo {year} {2021})},\ \Eprint
  {http://arxiv.org/abs/2007.02463} {arXiv:2007.02463 [nucl-th]} \BibitemShut
  {NoStop}%
\bibitem [{\citenamefont {Landau}\ and\ \citenamefont
  {Lifshitz}(1980)}]{StatLandau}%
  \BibitemOpen
  \bibfield  {author} {\bibinfo {author} {\bibfnamefont {L.~D.}\ \bibnamefont
  {Landau}}\ and\ \bibinfo {author} {\bibfnamefont {E.~M.}\ \bibnamefont
  {Lifshitz}},\ }\href@noop {} {\emph {\bibinfo {title} {Statistical
  Physics}}}\ (\bibinfo  {publisher} {Pergamon Press},\ \bibinfo {year}
  {1980})\BibitemShut {NoStop}%
\bibitem [{\citenamefont {Vovchenko}\ and\ \citenamefont
  {Koch}(2022{\natexlab{a}})}]{Vovchenko:2022szk}%
  \BibitemOpen
  \bibfield  {author} {\bibinfo {author} {\bibfnamefont {V.}~\bibnamefont
  {Vovchenko}}\ and\ \bibinfo {author} {\bibfnamefont {V.}~\bibnamefont
  {Koch}},\ }\href {\doibase 10.1016/j.physletb.2022.137368} {\bibfield
  {journal} {\bibinfo  {journal} {Phys. Lett. B}\ }\textbf {\bibinfo {volume}
  {833}},\ \bibinfo {pages} {137368} (\bibinfo {year} {2022}{\natexlab{a}})},\
  \Eprint {http://arxiv.org/abs/2204.00137} {arXiv:2204.00137 [hep-ph]}
  \BibitemShut {NoStop}%
\bibitem [{\citenamefont {Poberezhnyuk}\ \emph {et~al.}(2023)\citenamefont
  {Poberezhnyuk}, \citenamefont {Vovchenko}, \citenamefont {Savchuk},
  \citenamefont {Koch}, \citenamefont {Gorenstein},\ and\ \citenamefont
  {Stoecker}}]{Poberezhnyuk:2022blc}%
  \BibitemOpen
  \bibfield  {author} {\bibinfo {author} {\bibfnamefont {R.~V.}\ \bibnamefont
  {Poberezhnyuk}}, \bibinfo {author} {\bibfnamefont {V.}~\bibnamefont
  {Vovchenko}}, \bibinfo {author} {\bibfnamefont {O.}~\bibnamefont {Savchuk}},
  \bibinfo {author} {\bibfnamefont {V.}~\bibnamefont {Koch}}, \bibinfo {author}
  {\bibfnamefont {M.~I.}\ \bibnamefont {Gorenstein}}, \ and\ \bibinfo {author}
  {\bibfnamefont {H.}~\bibnamefont {Stoecker}},\ }\href {\doibase
  10.1051/epjconf/202327601005} {\bibfield  {journal} {\bibinfo  {journal} {EPJ
  Web Conf.}\ }\textbf {\bibinfo {volume} {276}},\ \bibinfo {pages} {01005}
  (\bibinfo {year} {2023})},\ \Eprint {http://arxiv.org/abs/2210.02960}
  {arXiv:2210.02960 [hep-ph]} \BibitemShut {NoStop}%
\bibitem [{\citenamefont {Braun-Munzinger}\ \emph
  {et~al.}(2019{\natexlab{b}})\citenamefont {Braun-Munzinger}, \citenamefont
  {Rustamov},\ and\ \citenamefont {Stachel}}]{Braun-Munzinger:2019yxj}%
  \BibitemOpen
  \bibfield  {author} {\bibinfo {author} {\bibfnamefont {P.}~\bibnamefont
  {Braun-Munzinger}}, \bibinfo {author} {\bibfnamefont {A.}~\bibnamefont
  {Rustamov}}, \ and\ \bibinfo {author} {\bibfnamefont {J.}~\bibnamefont
  {Stachel}},\ }\href@noop {} {\  (\bibinfo {year} {2019}{\natexlab{b}})},\
  \Eprint {http://arxiv.org/abs/1907.03032} {arXiv:1907.03032 [nucl-th]}
  \BibitemShut {NoStop}%
\bibitem [{\citenamefont {Redlich}\ and\ \citenamefont
  {Turko}(1980)}]{Redlich:1979bf}%
  \BibitemOpen
  \bibfield  {author} {\bibinfo {author} {\bibfnamefont {K.}~\bibnamefont
  {Redlich}}\ and\ \bibinfo {author} {\bibfnamefont {L.}~\bibnamefont
  {Turko}},\ }\href {\doibase 10.1007/BF01421776} {\bibfield  {journal}
  {\bibinfo  {journal} {Z. Phys. C}\ }\textbf {\bibinfo {volume} {5}},\
  \bibinfo {pages} {201} (\bibinfo {year} {1980})}\BibitemShut {NoStop}%
\bibitem [{\citenamefont {Hagedorn}\ and\ \citenamefont
  {Redlich}(1985)}]{Hagedorn:1984uy}%
  \BibitemOpen
  \bibfield  {author} {\bibinfo {author} {\bibfnamefont {R.}~\bibnamefont
  {Hagedorn}}\ and\ \bibinfo {author} {\bibfnamefont {K.}~\bibnamefont
  {Redlich}},\ }\href {\doibase 10.1007/BF01436508} {\bibfield  {journal}
  {\bibinfo  {journal} {Z. Phys. C}\ }\textbf {\bibinfo {volume} {27}},\
  \bibinfo {pages} {541} (\bibinfo {year} {1985})}\BibitemShut {NoStop}%
\bibitem [{\citenamefont {Braun-Munzinger}\ \emph {et~al.}(2003)\citenamefont
  {Braun-Munzinger}, \citenamefont {Redlich},\ and\ \citenamefont
  {Stachel}}]{Braun-Munzinger:2003pwq}%
  \BibitemOpen
  \bibfield  {author} {\bibinfo {author} {\bibfnamefont {P.}~\bibnamefont
  {Braun-Munzinger}}, \bibinfo {author} {\bibfnamefont {K.}~\bibnamefont
  {Redlich}}, \ and\ \bibinfo {author} {\bibfnamefont {J.}~\bibnamefont
  {Stachel}},\ }\bibfield  {booktitle} {\emph {\bibinfo {booktitle}
  {{Quark-Gluon Plasma 2, Edited By: Rudolph C. Hwa and Xin-Nian Wang }}},\
  }\href {\doibase 10.1142/9789812795533_0008} {\bibfield  {journal} {\bibinfo
  {journal} {World Scientific}\ ,\ \bibinfo {pages} {491}} (\bibinfo {year}
  {2003})},\ \Eprint {http://arxiv.org/abs/nucl-th/0304013}
  {arXiv:nucl-th/0304013} \BibitemShut {NoStop}%
\bibitem [{\citenamefont {Begun}\ \emph {et~al.}(2005)\citenamefont {Begun},
  \citenamefont {Gorenstein},\ and\ \citenamefont {Zozulya}}]{Begun:2004zb}%
  \BibitemOpen
  \bibfield  {author} {\bibinfo {author} {\bibfnamefont {V.~V.}\ \bibnamefont
  {Begun}}, \bibinfo {author} {\bibfnamefont {M.~I.}\ \bibnamefont
  {Gorenstein}}, \ and\ \bibinfo {author} {\bibfnamefont {O.~S.}\ \bibnamefont
  {Zozulya}},\ }\href {\doibase 10.1103/PhysRevC.72.014902} {\bibfield
  {journal} {\bibinfo  {journal} {Phys. Rev. C}\ }\textbf {\bibinfo {volume}
  {72}},\ \bibinfo {pages} {014902} (\bibinfo {year} {2005})},\ \Eprint
  {http://arxiv.org/abs/nucl-th/0411003} {arXiv:nucl-th/0411003} \BibitemShut
  {NoStop}%
\bibitem [{\citenamefont {Becattini}(1996)}]{Becattini:1995if}%
  \BibitemOpen
  \bibfield  {author} {\bibinfo {author} {\bibfnamefont {F.}~\bibnamefont
  {Becattini}},\ }\href {\doibase 10.1007/BF02907431} {\bibfield  {journal}
  {\bibinfo  {journal} {Z. Phys. C}\ }\textbf {\bibinfo {volume} {69}},\
  \bibinfo {pages} {485} (\bibinfo {year} {1996})}\BibitemShut {NoStop}%
\bibitem [{\citenamefont {{M. Abramowitz, and I. A. Stegun
  (Eds.)}}(1972)}]{Abramowitz:1972}%
  \BibitemOpen
  \bibfield  {author} {\bibinfo {author} {\bibnamefont {{M. Abramowitz, and I.
  A. Stegun (Eds.)}}},\ }\href@noop {} {\emph {\bibinfo {title} {{Handbook of
  Mathematical Functions with Formulas, Graphs, and Mathematical Tables}}}}\
  (\bibinfo  {publisher} {Dover Publications Inc., New York},\ \bibinfo {year}
  {1972})\BibitemShut {NoStop}%
\bibitem [{\citenamefont {Cleymans}\ and\ \citenamefont
  {Muronga}(1996)}]{Cleymans:1996yg}%
  \BibitemOpen
  \bibfield  {author} {\bibinfo {author} {\bibfnamefont {J.}~\bibnamefont
  {Cleymans}}\ and\ \bibinfo {author} {\bibfnamefont {A.}~\bibnamefont
  {Muronga}},\ }\href {\doibase 10.1016/0370-2693(96)01135-5} {\bibfield
  {journal} {\bibinfo  {journal} {Phys. Lett. B}\ }\textbf {\bibinfo {volume}
  {388}},\ \bibinfo {pages} {5} (\bibinfo {year} {1996})},\ \Eprint
  {http://arxiv.org/abs/nucl-th/9607042} {arXiv:nucl-th/9607042} \BibitemShut
  {NoStop}%
\bibitem [{\citenamefont {Cleymans}\ \emph {et~al.}(1991)\citenamefont
  {Cleymans}, \citenamefont {Redlich},\ and\ \citenamefont
  {Suhonen}}]{Cleymans:1990mn}%
  \BibitemOpen
  \bibfield  {author} {\bibinfo {author} {\bibfnamefont {J.}~\bibnamefont
  {Cleymans}}, \bibinfo {author} {\bibfnamefont {K.}~\bibnamefont {Redlich}}, \
  and\ \bibinfo {author} {\bibfnamefont {E.}~\bibnamefont {Suhonen}},\ }\href
  {\doibase 10.1007/BF01579571} {\bibfield  {journal} {\bibinfo  {journal} {Z.
  Phys. C}\ }\textbf {\bibinfo {volume} {51}},\ \bibinfo {pages} {137}
  (\bibinfo {year} {1991})}\BibitemShut {NoStop}%
\bibitem [{\citenamefont {Hamieh}\ \emph {et~al.}(2000)\citenamefont {Hamieh},
  \citenamefont {Redlich},\ and\ \citenamefont {Tounsi}}]{Hamieh:2000tk}%
  \BibitemOpen
  \bibfield  {author} {\bibinfo {author} {\bibfnamefont {S.}~\bibnamefont
  {Hamieh}}, \bibinfo {author} {\bibfnamefont {K.}~\bibnamefont {Redlich}}, \
  and\ \bibinfo {author} {\bibfnamefont {A.}~\bibnamefont {Tounsi}},\ }\href
  {\doibase 10.1016/S0370-2693(00)00762-0} {\bibfield  {journal} {\bibinfo
  {journal} {Phys. Lett. B}\ }\textbf {\bibinfo {volume} {486}},\ \bibinfo
  {pages} {61} (\bibinfo {year} {2000})},\ \Eprint
  {http://arxiv.org/abs/hep-ph/0006024} {arXiv:hep-ph/0006024} \BibitemShut
  {NoStop}%
\bibitem [{\citenamefont {Cleymans}\ \emph {et~al.}(1999)\citenamefont
  {Cleymans}, \citenamefont {Oeschler},\ and\ \citenamefont
  {Redlich}}]{Cleymans:1998yb}%
  \BibitemOpen
  \bibfield  {author} {\bibinfo {author} {\bibfnamefont {J.}~\bibnamefont
  {Cleymans}}, \bibinfo {author} {\bibfnamefont {H.}~\bibnamefont {Oeschler}},
  \ and\ \bibinfo {author} {\bibfnamefont {K.}~\bibnamefont {Redlich}},\ }\href
  {\doibase 10.1103/PhysRevC.59.1663} {\bibfield  {journal} {\bibinfo
  {journal} {Phys. Rev. C}\ }\textbf {\bibinfo {volume} {59}},\ \bibinfo
  {pages} {1663} (\bibinfo {year} {1999})},\ \Eprint
  {http://arxiv.org/abs/nucl-th/9809027} {arXiv:nucl-th/9809027} \BibitemShut
  {NoStop}%
\bibitem [{\citenamefont {Tounsi}\ and\ \citenamefont
  {Redlich}(2001)}]{Tounsi:2001ck}%
  \BibitemOpen
  \bibfield  {author} {\bibinfo {author} {\bibfnamefont {A.}~\bibnamefont
  {Tounsi}}\ and\ \bibinfo {author} {\bibfnamefont {K.}~\bibnamefont
  {Redlich}},\ }\href@noop {} {\  (\bibinfo {year} {2001})},\ \Eprint
  {http://arxiv.org/abs/hep-ph/0111159} {arXiv:hep-ph/0111159} \BibitemShut
  {NoStop}%
\bibitem [{\citenamefont {Sharma}\ \emph {et~al.}(2023)\citenamefont {Sharma},
  \citenamefont {Kumar}, \citenamefont {Lo},\ and\ \citenamefont
  {Redlich}}]{Sharma:2022poi}%
  \BibitemOpen
  \bibfield  {author} {\bibinfo {author} {\bibfnamefont {N.}~\bibnamefont
  {Sharma}}, \bibinfo {author} {\bibfnamefont {L.}~\bibnamefont {Kumar}},
  \bibinfo {author} {\bibfnamefont {P.~M.}\ \bibnamefont {Lo}}, \ and\ \bibinfo
  {author} {\bibfnamefont {K.}~\bibnamefont {Redlich}},\ }\href {\doibase
  10.1103/PhysRevC.107.054903} {\bibfield  {journal} {\bibinfo  {journal}
  {Phys. Rev. C}\ }\textbf {\bibinfo {volume} {107}},\ \bibinfo {pages}
  {054903} (\bibinfo {year} {2023})},\ \Eprint
  {http://arxiv.org/abs/2210.15617} {arXiv:2210.15617 [nucl-th]} \BibitemShut
  {NoStop}%
\bibitem [{\citenamefont {Arsene}\ \emph {et~al.}(2009)\citenamefont {Arsene}
  \emph {et~al.}}]{BRAHMS:2009wlg}%
  \BibitemOpen
  \bibfield  {author} {\bibinfo {author} {\bibfnamefont {I.~C.}\ \bibnamefont
  {Arsene}} \emph {et~al.} (\bibinfo {collaboration} {BRAHMS}),\ }\href
  {\doibase 10.1016/j.physletb.2009.05.049} {\bibfield  {journal} {\bibinfo
  {journal} {Phys. Lett. B}\ }\textbf {\bibinfo {volume} {677}},\ \bibinfo
  {pages} {267} (\bibinfo {year} {2009})},\ \Eprint
  {http://arxiv.org/abs/0901.0872} {arXiv:0901.0872 [nucl-ex]} \BibitemShut
  {NoStop}%
\bibitem [{\citenamefont {Adam}\ \emph {et~al.}(2017)\citenamefont {Adam} \emph
  {et~al.}}]{ALICE:2016fbt}%
  \BibitemOpen
  \bibfield  {author} {\bibinfo {author} {\bibfnamefont {J.}~\bibnamefont
  {Adam}} \emph {et~al.} (\bibinfo {collaboration} {ALICE}),\ }\href {\doibase
  10.1016/j.physletb.2017.07.017} {\bibfield  {journal} {\bibinfo  {journal}
  {Phys. Lett. B}\ }\textbf {\bibinfo {volume} {772}},\ \bibinfo {pages} {567}
  (\bibinfo {year} {2017})},\ \Eprint {http://arxiv.org/abs/1612.08966}
  {arXiv:1612.08966 [nucl-ex]} \BibitemShut {NoStop}%
\bibitem [{\citenamefont {Acharya}\ \emph
  {et~al.}(2023{\natexlab{b}})\citenamefont {Acharya} \emph
  {et~al.}}]{ALICE:2023ulv}%
  \BibitemOpen
  \bibfield  {author} {\bibinfo {author} {\bibfnamefont {S.}~\bibnamefont
  {Acharya}} \emph {et~al.} (\bibinfo {collaboration} {ALICE}),\ }\href@noop {}
  {\  (\bibinfo {year} {2023}{\natexlab{b}})},\ \Eprint
  {http://arxiv.org/abs/2311.13332} {arXiv:2311.13332 [nucl-ex]} \BibitemShut
  {NoStop}%
\bibitem [{\citenamefont {Braun-Munzinger}\ \emph {et~al.}(2015)\citenamefont
  {Braun-Munzinger}, \citenamefont {Kalweit}, \citenamefont {Redlich},\ and\
  \citenamefont {Stachel}}]{Braun-Munzinger:2014lba}%
  \BibitemOpen
  \bibfield  {author} {\bibinfo {author} {\bibfnamefont {P.}~\bibnamefont
  {Braun-Munzinger}}, \bibinfo {author} {\bibfnamefont {A.}~\bibnamefont
  {Kalweit}}, \bibinfo {author} {\bibfnamefont {K.}~\bibnamefont {Redlich}}, \
  and\ \bibinfo {author} {\bibfnamefont {J.}~\bibnamefont {Stachel}},\ }\href
  {\doibase 10.1016/j.physletb.2015.05.077} {\bibfield  {journal} {\bibinfo
  {journal} {Phys. Lett. B}\ }\textbf {\bibinfo {volume} {747}},\ \bibinfo
  {pages} {292} (\bibinfo {year} {2015})},\ \Eprint
  {http://arxiv.org/abs/1412.8614} {arXiv:1412.8614 [hep-ph]} \BibitemShut
  {NoStop}%
\bibitem [{\citenamefont {Aamodt}\ \emph {et~al.}(2011)\citenamefont {Aamodt}
  \emph {et~al.}}]{ALICE:2011dyt}%
  \BibitemOpen
  \bibfield  {author} {\bibinfo {author} {\bibfnamefont {K.}~\bibnamefont
  {Aamodt}} \emph {et~al.} (\bibinfo {collaboration} {ALICE}),\ }\href
  {\doibase 10.1016/j.physletb.2010.12.053} {\bibfield  {journal} {\bibinfo
  {journal} {Phys. Lett. B}\ }\textbf {\bibinfo {volume} {696}},\ \bibinfo
  {pages} {328} (\bibinfo {year} {2011})},\ \Eprint
  {http://arxiv.org/abs/1012.4035} {arXiv:1012.4035 [nucl-ex]} \BibitemShut
  {NoStop}%
\bibitem [{\citenamefont {{Golub, G.H., Van Loan,
  C.F.}}(1996)}]{Cholesky:1996ab}%
  \BibitemOpen
  \bibfield  {author} {\bibinfo {author} {\bibnamefont {{Golub, G.H., Van Loan,
  C.F.}}},\ }\href@noop {} {\emph {\bibinfo {title} {{Matrix computations}}}}\
  (\bibinfo  {publisher} {Johns Hopkins University Press},\ \bibinfo {year}
  {1996})\BibitemShut {NoStop}%
\bibitem [{\citenamefont {Barrette}\ \emph {et~al.}(2000)\citenamefont
  {Barrette} \emph {et~al.}}]{E877:1999qdc}%
  \BibitemOpen
  \bibfield  {author} {\bibinfo {author} {\bibfnamefont {J.}~\bibnamefont
  {Barrette}} \emph {et~al.} (\bibinfo {collaboration} {E877}),\ }\href
  {\doibase 10.1103/PhysRevC.62.024901} {\bibfield  {journal} {\bibinfo
  {journal} {Phys. Rev. C}\ }\textbf {\bibinfo {volume} {62}},\ \bibinfo
  {pages} {024901} (\bibinfo {year} {2000})},\ \Eprint
  {http://arxiv.org/abs/nucl-ex/9910004} {arXiv:nucl-ex/9910004} \BibitemShut
  {NoStop}%
\bibitem [{\citenamefont {Ahle}\ \emph {et~al.}(1998)\citenamefont {Ahle} \emph
  {et~al.}}]{E-802:1998xum}%
  \BibitemOpen
  \bibfield  {author} {\bibinfo {author} {\bibfnamefont {L.}~\bibnamefont
  {Ahle}} \emph {et~al.} (\bibinfo {collaboration} {E-802}),\ }\href {\doibase
  10.1103/PhysRevC.57.R466} {\bibfield  {journal} {\bibinfo  {journal} {Phys.
  Rev. C}\ }\textbf {\bibinfo {volume} {57}},\ \bibinfo {pages} {R466}
  (\bibinfo {year} {1998})}\BibitemShut {NoStop}%
\bibitem [{\citenamefont {Metropolis}\ \emph {et~al.}(1953)\citenamefont
  {Metropolis}, \citenamefont {Rosenbluth}, \citenamefont {Rosenbluth},
  \citenamefont {Teller},\ and\ \citenamefont {Teller}}]{Metropolis:1953am}%
  \BibitemOpen
  \bibfield  {author} {\bibinfo {author} {\bibfnamefont {N.}~\bibnamefont
  {Metropolis}}, \bibinfo {author} {\bibfnamefont {A.~W.}\ \bibnamefont
  {Rosenbluth}}, \bibinfo {author} {\bibfnamefont {M.~N.}\ \bibnamefont
  {Rosenbluth}}, \bibinfo {author} {\bibfnamefont {A.~H.}\ \bibnamefont
  {Teller}}, \ and\ \bibinfo {author} {\bibfnamefont {E.}~\bibnamefont
  {Teller}},\ }\href {\doibase 10.1063/1.1699114} {\bibfield  {journal}
  {\bibinfo  {journal} {J. Chem. Phys.}\ }\textbf {\bibinfo {volume} {21}},\
  \bibinfo {pages} {1087} (\bibinfo {year} {1953})}\BibitemShut {NoStop}%
\bibitem [{\citenamefont {Charmpis}\ and\ \citenamefont
  {Panteli}(2004)}]{Charmpis:2004}%
  \BibitemOpen
  \bibfield  {author} {\bibinfo {author} {\bibfnamefont {D.~C.}\ \bibnamefont
  {Charmpis}}\ and\ \bibinfo {author} {\bibfnamefont {P.~L.}\ \bibnamefont
  {Panteli}},\ }\href@noop {} {\bibfield  {journal} {\bibinfo  {journal} {Comp.
  Stat.}\ }\textbf {\bibinfo {volume} {19}},\ \bibinfo {pages} {283} (\bibinfo
  {year} {2004})}\BibitemShut {NoStop}%
\bibitem [{\citenamefont {Kirkpatrick}\ \emph {et~al.}(1983)\citenamefont
  {Kirkpatrick}, \citenamefont {Gelatt},\ and\ \citenamefont
  {Vecchi}}]{Kirkpatrick:1983zz}%
  \BibitemOpen
  \bibfield  {author} {\bibinfo {author} {\bibfnamefont {S.}~\bibnamefont
  {Kirkpatrick}}, \bibinfo {author} {\bibfnamefont {C.~D.}\ \bibnamefont
  {Gelatt}}, \ and\ \bibinfo {author} {\bibfnamefont {M.~P.}\ \bibnamefont
  {Vecchi}},\ }\href {\doibase 10.1126/science.220.4598.671} {\bibfield
  {journal} {\bibinfo  {journal} {Science}\ }\textbf {\bibinfo {volume}
  {220}},\ \bibinfo {pages} {671} (\bibinfo {year} {1983})}\BibitemShut
  {NoStop}%
\bibitem [{\citenamefont {Sasaki}\ \emph {et~al.}(2007)\citenamefont {Sasaki},
  \citenamefont {Friman},\ and\ \citenamefont {Redlich}}]{Sasaki:2007db}%
  \BibitemOpen
  \bibfield  {author} {\bibinfo {author} {\bibfnamefont {C.}~\bibnamefont
  {Sasaki}}, \bibinfo {author} {\bibfnamefont {B.}~\bibnamefont {Friman}}, \
  and\ \bibinfo {author} {\bibfnamefont {K.}~\bibnamefont {Redlich}},\ }\href
  {\doibase 10.1103/PhysRevLett.99.232301} {\bibfield  {journal} {\bibinfo
  {journal} {Phys. Rev. Lett.}\ }\textbf {\bibinfo {volume} {99}},\ \bibinfo
  {pages} {232301} (\bibinfo {year} {2007})},\ \Eprint
  {http://arxiv.org/abs/hep-ph/0702254} {arXiv:hep-ph/0702254} \BibitemShut
  {NoStop}%
\bibitem [{\citenamefont {Gyulassy}\ and\ \citenamefont
  {Wang}(1994)}]{Gyulassy:1994ew}%
  \BibitemOpen
  \bibfield  {author} {\bibinfo {author} {\bibfnamefont {M.}~\bibnamefont
  {Gyulassy}}\ and\ \bibinfo {author} {\bibfnamefont {X.-N.}\ \bibnamefont
  {Wang}},\ }\href {\doibase 10.1016/0010-4655(94)90057-4} {\bibfield
  {journal} {\bibinfo  {journal} {Comput. Phys. Commun.}\ }\textbf {\bibinfo
  {volume} {83}},\ \bibinfo {pages} {307} (\bibinfo {year} {1994})},\ \Eprint
  {http://arxiv.org/abs/nucl-th/9502021} {arXiv:nucl-th/9502021} \BibitemShut
  {NoStop}%
\bibitem [{\citenamefont {Andersson}\ \emph {et~al.}(1983)\citenamefont
  {Andersson}, \citenamefont {Gustafson}, \citenamefont {Ingelman},\ and\
  \citenamefont {Sjostrand}}]{Andersson:1983ia}%
  \BibitemOpen
  \bibfield  {author} {\bibinfo {author} {\bibfnamefont {B.}~\bibnamefont
  {Andersson}}, \bibinfo {author} {\bibfnamefont {G.}~\bibnamefont
  {Gustafson}}, \bibinfo {author} {\bibfnamefont {G.}~\bibnamefont {Ingelman}},
  \ and\ \bibinfo {author} {\bibfnamefont {T.}~\bibnamefont {Sjostrand}},\
  }\href {\doibase 10.1016/0370-1573(83)90080-7} {\bibfield  {journal}
  {\bibinfo  {journal} {Phys. Rept.}\ }\textbf {\bibinfo {volume} {97}},\
  \bibinfo {pages} {31} (\bibinfo {year} {1983})}\BibitemShut {NoStop}%
\bibitem [{\citenamefont {Eden}\ and\ \citenamefont
  {Gustafson}(1997)}]{Eden:1996xi}%
  \BibitemOpen
  \bibfield  {author} {\bibinfo {author} {\bibfnamefont {P.}~\bibnamefont
  {Eden}}\ and\ \bibinfo {author} {\bibfnamefont {G.}~\bibnamefont
  {Gustafson}},\ }\href {\doibase 10.1007/s002880050445} {\bibfield  {journal}
  {\bibinfo  {journal} {Z. Phys. C}\ }\textbf {\bibinfo {volume} {75}},\
  \bibinfo {pages} {41} (\bibinfo {year} {1997})},\ \Eprint
  {http://arxiv.org/abs/hep-ph/9606454} {arXiv:hep-ph/9606454} \BibitemShut
  {NoStop}%
\bibitem [{\citenamefont {Savchuk}\ \emph {et~al.}(2022)\citenamefont
  {Savchuk}, \citenamefont {Vovchenko}, \citenamefont {Koch}, \citenamefont
  {Steinheimer},\ and\ \citenamefont {Stoecker}}]{Savchuk:2021aog}%
  \BibitemOpen
  \bibfield  {author} {\bibinfo {author} {\bibfnamefont {O.}~\bibnamefont
  {Savchuk}}, \bibinfo {author} {\bibfnamefont {V.}~\bibnamefont {Vovchenko}},
  \bibinfo {author} {\bibfnamefont {V.}~\bibnamefont {Koch}}, \bibinfo {author}
  {\bibfnamefont {J.}~\bibnamefont {Steinheimer}}, \ and\ \bibinfo {author}
  {\bibfnamefont {H.}~\bibnamefont {Stoecker}},\ }\href {\doibase
  10.1016/j.physletb.2022.136983} {\bibfield  {journal} {\bibinfo  {journal}
  {Phys. Lett. B}\ }\textbf {\bibinfo {volume} {827}},\ \bibinfo {pages}
  {136983} (\bibinfo {year} {2022})},\ \Eprint
  {http://arxiv.org/abs/2106.08239} {arXiv:2106.08239 [hep-ph]} \BibitemShut
  {NoStop}%
\bibitem [{\citenamefont {Vovchenko}\ and\ \citenamefont
  {Koch}(2022{\natexlab{b}})}]{Vovchenko:2022xil}%
  \BibitemOpen
  \bibfield  {author} {\bibinfo {author} {\bibfnamefont {V.}~\bibnamefont
  {Vovchenko}}\ and\ \bibinfo {author} {\bibfnamefont {V.}~\bibnamefont
  {Koch}},\ }\href {\doibase 10.1016/j.physletb.2022.137577} {\bibfield
  {journal} {\bibinfo  {journal} {Phys. Lett. B}\ }\textbf {\bibinfo {volume}
  {835}},\ \bibinfo {pages} {137577} (\bibinfo {year} {2022}{\natexlab{b}})},\
  \Eprint {http://arxiv.org/abs/2210.15641} {arXiv:2210.15641 [nucl-th]}
  \BibitemShut {NoStop}%
\bibitem [{\citenamefont {Kitazawa}(2015)}]{Kitazawa:2015ira}%
  \BibitemOpen
  \bibfield  {author} {\bibinfo {author} {\bibfnamefont {M.}~\bibnamefont
  {Kitazawa}},\ }\href {\doibase 10.1016/j.nuclphysa.2015.07.008} {\bibfield
  {journal} {\bibinfo  {journal} {Nucl. Phys. A}\ }\textbf {\bibinfo {volume}
  {942}},\ \bibinfo {pages} {65} (\bibinfo {year} {2015})},\ \Eprint
  {http://arxiv.org/abs/1505.04349} {arXiv:1505.04349 [nucl-th]} \BibitemShut
  {NoStop}%
\bibitem [{\citenamefont {Sakaida}\ \emph {et~al.}(2014)\citenamefont
  {Sakaida}, \citenamefont {Asakawa},\ and\ \citenamefont
  {Kitazawa}}]{Sakaida:2014pya}%
  \BibitemOpen
  \bibfield  {author} {\bibinfo {author} {\bibfnamefont {M.}~\bibnamefont
  {Sakaida}}, \bibinfo {author} {\bibfnamefont {M.}~\bibnamefont {Asakawa}}, \
  and\ \bibinfo {author} {\bibfnamefont {M.}~\bibnamefont {Kitazawa}},\ }\href
  {\doibase 10.1103/PhysRevC.90.064911} {\bibfield  {journal} {\bibinfo
  {journal} {Phys. Rev. C}\ }\textbf {\bibinfo {volume} {90}},\ \bibinfo
  {pages} {064911} (\bibinfo {year} {2014})},\ \Eprint
  {http://arxiv.org/abs/1409.6866} {arXiv:1409.6866 [nucl-th]} \BibitemShut
  {NoStop}%
\end{thebibliography}%

\end{document}